\documentclass[astronomy,article,submit,moreauthors]{Definitions/mdpi}
\AtBeginDocument{\let\linenumbers\relax\nolinenumbers}

\firstpage{1}
\pubvolume{5}
\issuenum{1}
\articlenumber{0}
\pubyear{2026}
\copyrightyear{2026}
\datereceived{ }
\daterevised{ }
\dateaccepted{ }
\datepublished{ }

\usepackage{comment}
\usepackage{pifont}
\newenvironment{subfigure}[1]{\begin{minipage}{#1}}{\end{minipage}}

\makeatletter
\newcommand{\seedcontinuedfloat}{\caption@reset@continuedfloat{figure}}
\newcounter{appendixfig}
\newcommand{\appfigsection}{\setcounter{appendixfig}{0}}
\newcommand{\appfigreset}{\stepcounter{appendixfig}\renewcommand{\thefigure}{\thesection-\arabic{appendixfig}}}
\makeatother
\usetikzlibrary{arrows.meta, positioning, shapes.geometric, calc}

\definecolor{boxgray}{RGB}{225,225,225}
\definecolor{cosmoblue}{RGB}{222,229,241}
\definecolor{cosmobluetext}{RGB}{31,56,110}
\definecolor{noisepurple}{RGB}{229,220,236}
\definecolor{noisepurpletext}{RGB}{95,49,120}
\definecolor{gtgreen}{RGB}{223,236,207}
\definecolor{gigreen}{RGB}{249,224,196}
\definecolor{rsqgreen}{RGB}{231,241,225}
\definecolor{psiteal}{RGB}{209,236,229}
\definecolor{steparrow}{RGB}{150,60,40}

\Title{Weak lensing data beyond flexion}

\Author{Ben D. Normann $^{1,}$*\orcidA{}, Martin H. Enoksen $^{2}$, Nicol\`o Oreste Pinciroli Vago $^{3,4,5}$\orcidB{} and Hans Georg Schaathun $^{1}$\orcidC{}}

\AuthorNames{Ben D. Normann, Martin H. Enoksen, Nicol\`o Oreste Pinciroli Vago and Hans Georg Schaathun}

\address{%
$^{1}$ \quad Department of ICT and Natural Sciences, Norwegian University of Science and Technology, Norway\\
$^{2}$ \quad Department of Physics, Norwegian University of Science and Technology, Norway\\
$^{3}$ \quad Department of Electronics, Information and Bioengineering, Politecnico di Milano, via G. Ponzio 34, I-20133 Milan, Italy\\
$^{4}$ \quad INAF -- Osservatorio Astronomico di Roma, Via Frascati 33, I-00078 Monte Porzio Catone, Italy\\
$^{5}$ \quad Center for Astrophysics | Harvard \& Smithsonian, 60 Garden Street, Cambridge, MA 02138, USA}

\corres{Correspondence: ben.d.normann@ntnu.no}

\abstract{In weak lensing, shear and convergence are well studied and have provided a wealth of information about the distribution of matter in the universe. More recently, flexion -- the next term in the multipole expansion of the lensing potential, responsible for the characteristic banana-shaped distortion of lensed images -- has been shown to provide additional constraints on the mass distribution of lensing systems, particularly for galaxy clusters. Here we investigate whether weak-lensing information beyond flexion is similarly useful for lens-mass reconstruction. Using the roulette-amplitude formalism, we construct synthetic datasets for a single SIE lens and for SIE and SIS cluster lenses, with and without simulated Gaussian noise, and perform a least-squares reconstruction of the lens parameters using amplitudes truncated at third (flexion), fourth, and fifth order. As a proof of concept, we find that higher-order amplitudes do provide additional, useful information: the benefit is modest for a single lens, where flexion already constrains most parameters, but becomes substantial for cluster lenses, where the rotation and ellipticity are otherwise poorly recovered. The improvement is most pronounced in the presence of noise, where third-order reconstruction can become unreliable while higher-order amplitudes restore much of the lost accuracy. These results suggest that, as weak-lensing surveys grow in depth and precision, data beyond flexion could meaningfully improve mass reconstruction for the most complex and noisiest lensing systems, motivating further work using observational rather than synthetic data.}

\keyword{gravitational lensing; weak lensing; flexion; galaxy clusters; lens-mass reconstruction}

\begin{document}

\section{Introduction}
\label{sec:introduction}
Gravitational lensing~\citep{bok:SEF1999} is the effect of light rays being bent by gravitational fields. It has proven to be a promising tool for collecting data from the night sky, and divided into two regimes: strong lensing and weak lensing~\citep{bok:principlesOfGravLens18}. The former typically produces multiple images and arcs, and consequently requires only one event to be detected. The latter, on the other hand, has at least historically required statistical treatment to be detected. Recent years have, however, seen a surge of interest in combining the two regimes for lens-mass reconstruction. This is particularly interesting for cluster lenses, where both effects appear: strong-lensing effects are apparent near the center of the cluster, whereas weak-lensing events constrain the outskirts of the cluster. In such cases, a joint analysis typically constrains the cluster lens mass better than relying on either effect alone.

\begin{figure}[H]
    \centering
    \includegraphics[width=0.9\linewidth]{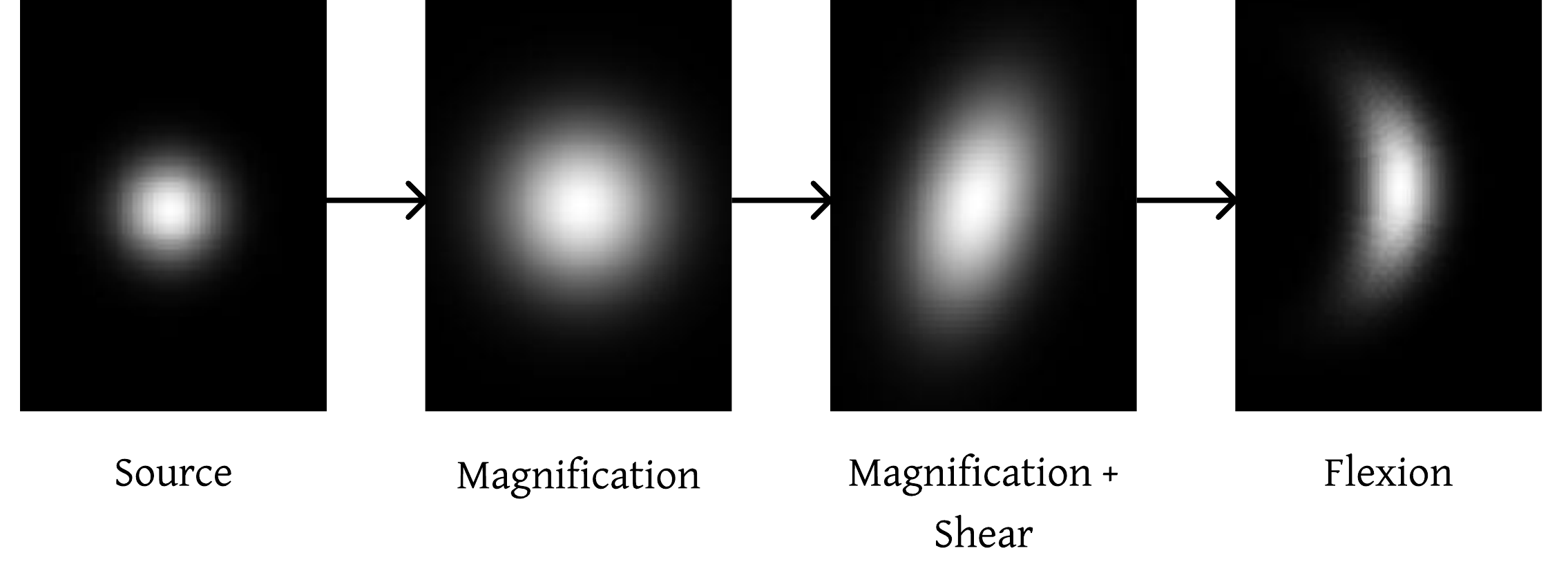}
    \caption{The effects of magnification, shear, and flexion on an initially circular source. Figure courtesy of the Master's thesis of Oda Hjemli~\citep{Hjemli2026}, generated using CosmoSim.\label{fig:magshearflexion}}
\end{figure}

Moreover, a joint lens-mass reconstruction procedure is actually possible, through so-called inverse approaches, like the maximum-likelihood approaches, as reviewed by~\cite{Normann2025}. In such studies one typically seeks to minimize a likelihood functional, where different kinds of constraints are added in a linear fashion. For lensing data specifically, this generically amounts to combining weak-lensing data with strong-lensing data. On the one hand, strong-lensing data typically consists of multiple images, or arclets. Weak-lensing data, on the other hand, consists of amplification, shear and so-called flexion effects. We will explain this in more detail in the next section. Suffice it here to say that amplification and shearing of a source arise as a second-order effect when Taylor-expanding the gravitational potential around the image. Flexion, on the other hand, causes the typical banana-shape\footnote{Or is it a jelly bean?} of a lensed image, and corresponds to a third-order term in the Taylor expansion. Figure~\ref{fig:magshearflexion} illustrates this progression, from an undistorted circular source through magnification, shear, and flexion~\citep{Cain2011,Leonard2007,bacon2010,Er2010,Cain2016,lanusse2016}. These 3rd-order terms have actually proven quite useful in probing the triaxial structure of cluster lenses~\citep{Hawken2009,Leonard2007,Irwin2006}.

This leads us to investigate the usefulness of data beyond flexion terms, as summarised by the following research question:

\noindent \textbf{Research question: }\textit{Does weak-lensing data beyond flexion terms improve lens-mass reconstruction?}

The answer to our research question is by no means obvious. Higher-order terms quickly decrease in magnitude, and it is hard to see that they will have detectable characteristic features distinct from flexion (the banana shape). On the other hand, as we shall see, higher-order terms grow in number, so even if their individual contribution is small, their collective contribution could potentially matter. Understanding the higher-order effects beyond flexion could thus prove useful in meeting the demands of the high-quality datasets of tomorrow, such as the forecasted collection of $100\,000$ strong-lensing images by Euclid alone~\citep{Lines2025}.

We have previously reported on the simulator we built for simulating terms beyond flexion in the so-called roulette formalism (CosmoSim; \citealt{hgs2023ecms}), along with some preliminary machine-learning results~\citep{hgs2023nik}. A paper describing the software release is currently in preparation, alongside a separate paper applying CosmoSim to real data~\citep{oda2026ngc6505}. The present paper extends previous work by focusing on a pipeline for applying CosmoSim to extract data beyond flexion.

The current paper thus extends previous analysis by the following contributions:
\begin{enumerate}[itemsep=0pt, parsep=0pt, topsep=0pt]
    \item \textit{Doing actual lens-parameter ($\boldsymbol{\lambda}$) reconstruction from data beyond flexion terms}. To the best of knowledge this has not been done before, and is as such the main contribution of this paper.
    \item \textit{Investigating cluster lenses}.
\end{enumerate}
Since \textbf{the main objective of this paper} is to assess to what extent higher-order terms are interesting \textit{in principle}, we still use synthetic data, with simulated noise, referring the application of CosmoSim to real data to the aforementioned~\citep{oda2026ngc6505}.

As we shall see, this work allows us to answer our research question in the affirmative: the results indicate that higher-order terms are beneficial for lens-mass reconstruction of cluster lenses.

\paragraph{Outline of the rest of the paper}
The rest of the paper is structured as follows: In Sect.~\ref{sec:theory} we provide the necessary theoretical backbone needed to understand higher-order lensing. We then go on to explain our methodology in Sect.~\ref{sec:method}. In Sect.~\ref{sec:results} we provide our results, before we conclude and present future outlooks in Sect.~\ref{sec:conclusion}.

\section{Theoretical Background}
\label{sec:theory}

\begin{figure}[H]
    \centering
    \includegraphics[width=0.8\linewidth]{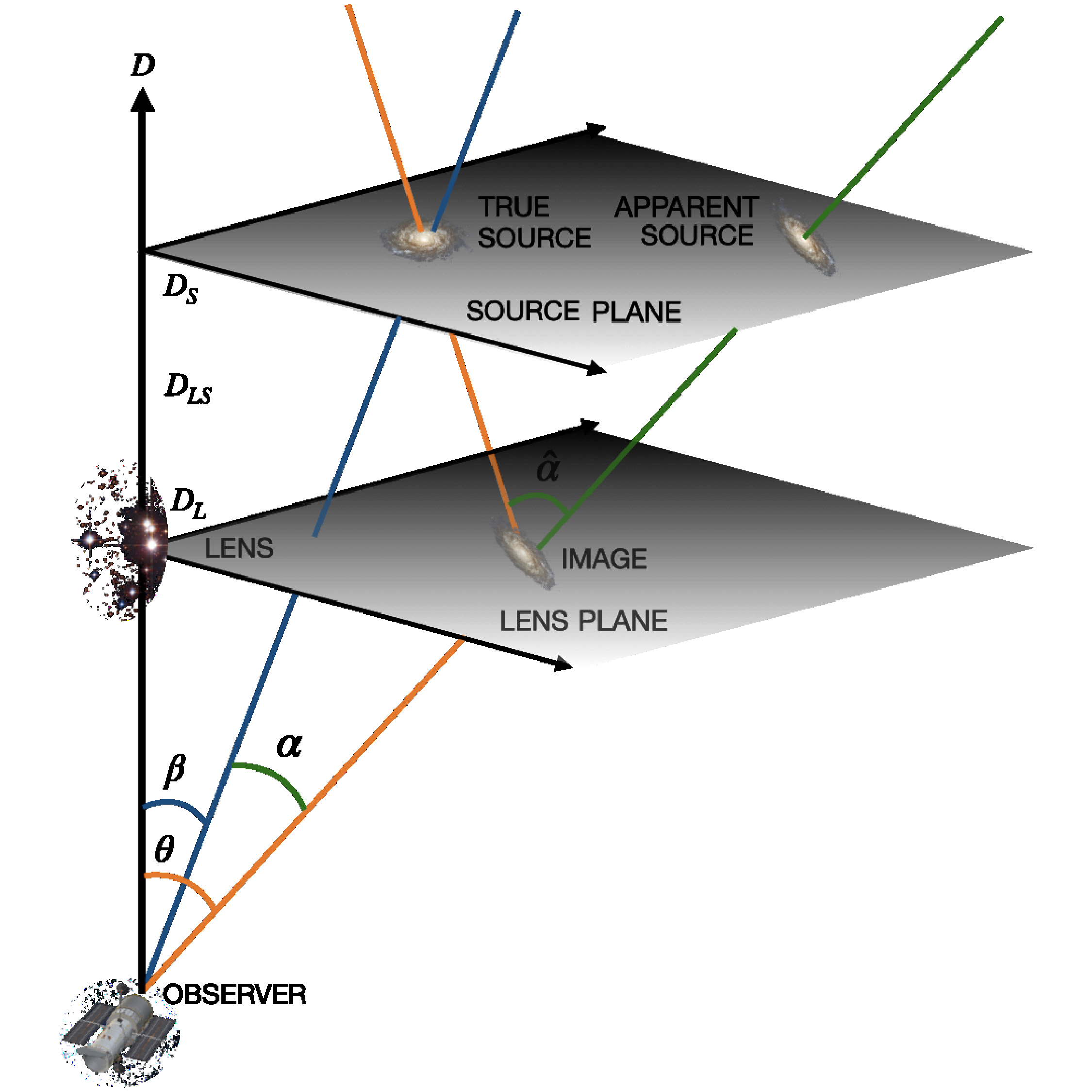}
    \caption{The lensing geometry. Here $D\textsubscript{L}$ and $D\textsubscript{S}$ are angular diameter distances from observer to lens and observer to source, whilst $D\textsubscript{LS}$ is the distance from lens to source. Furthermore, $\beta$ and $\theta$ are angular separations of source and image from the optical axis. Whilst $\alpha$ is the reduced deflection angle, the light-ray deflection is denoted by $\hat{\alpha}$. First used by the authors in~\citep{Normann2025}.\label{fig:lensing_geometry}}
\end{figure}

A typical starting point for gravitational lensing is the \textit{thin-lens} approximation~\citep{bok:SEF1999,kormann94b}, where the 3-dimensional gravitational potential $\Psi$ is integrated along the line of sight. The resulting two-dimensional gravitational potential $\psi(\boldsymbol{\theta})$ thus obeys Poisson's equation for gravity in the two-dimensional \textit{lens plane}\footnote{Locally the observer's sphere is a plane.} orthogonal to the observer and at a distance $D_{\mathrm{L}}$, as shown in Figure~\ref{fig:lensing_geometry}. That is, one has
\begin{equation}
    \label{poisson2D}
\nabla^2\psi=2\kappa\quad\quad\textrm{Poisson's equation},
\end{equation}
where $\nabla\equiv[\partial_1,\partial_2]\equiv[\frac{\partial}{\partial \theta_1},\frac{\partial}{\partial \theta_2}]$ is the nabla operator in angular coordinates of the lens plane; $\boldsymbol{\theta}=(\theta_1,\theta_2)$. Moreover, $\kappa$ is the \textit{dimensionless surface-mass density} (also known as \textit{convergence}). It is given by
\begin{equation}
    \kappa(\boldsymbol{\theta}) = \frac{\Sigma(\boldsymbol{\theta})}{\Sigma\textsubscript{cr}},
\end{equation}
where $\Sigma$ is the surface-mass density and $\Sigma_\textrm{cr}$ is a critical value defined such that
\begin{equation}
    \Sigma\textsubscript{cr} = \frac{c^2}{4\pi G}\frac{D\textsubscript{S}}{D\textsubscript{L}D\textsubscript{LS}}.
\end{equation}

One may use Green's functions to show that an integral solution to the above Poisson's equation is given by
\begin{equation}\label{lensing_potential}
\psi(\boldsymbol{\theta})=\frac{1}{\pi} \int_{\mathbb{R}^2} \mathrm{~d}^2 \theta^{\prime} \kappa\left(\boldsymbol{\theta}^{\prime}\right) \ln \left|\boldsymbol{\theta}-\boldsymbol{\theta}^{\prime}\right|.
\end{equation}
Furthermore, it follows that the so-called \textit{reduced deflection angle} is given by
\begin{equation}
    \label{redDeflAng}
    \boldsymbol{\alpha}=\nabla\psi.
\end{equation}
Consider now a beam of light emanating from a source at a distance $D_{\mathrm{S}}$, and let $\boldsymbol{\theta}=(\theta_1,\theta_2)$ be the observed angular position of the image in the lens plane. It is a simple matter of trigonometry to show that the angular position $\boldsymbol{\beta}=(\beta_1,\beta_2)$ of the source producing that image is given by
 \begin{equation}
 \label{LensMap}
     \boldsymbol{\beta} = f(\boldsymbol{\theta})=\boldsymbol{\theta}-\nabla\psi(\boldsymbol{\theta}),\quad\quad\textrm{ray-trace eq.},
 \end{equation}
If a source is placed directly behind ($\boldsymbol{\beta}=0$) a symmetric lens, the above equation shows that the image will form a ring. The radius of this ring is known as the (angular) Einstein radius, throughout this paper denoted by $\theta_\text{E}$, and used as a quantifier for the strength of a particular gravitational lens system.

If the gravitational field is more or less constant over the cross-section of the image, the ray-trace equation provides physically interesting results. Otherwise one must move from pixel to pixel in the source, mapping the light in it to the lens plane via the ray-trace mapping above.

\subsection{Shear and Flexion}
Alternatively, instead of ray-tracing pixel by pixel one could Taylor-expand around a point of choice, such as the image position $\boldsymbol{\theta}_0$ in the lens-plane. The following discussion will be most convenient to conduct in the complex plane. To that end, define
\begin{align}
 \label{complexQ}
\boldsymbol{\beta}_c=\beta_1+i\beta_2\quad\textrm{and}\quad \boldsymbol{\theta}_c=\theta_1+i\theta_2,
\end{align}
Furthermore, let the so-called \textit{reduced} counterpart (so as to take mass-sheet degeneracy into account) be denoted by a hat ($\hat{\,}$) and related to the original variable by
\begin{align}
    \hat{X}\equiv\frac{X}{1-\kappa}\quad\quad\textrm{(reduced var.)}
\end{align}
 To second order, the map between (complex) source-plane and lens-plane positions is now given by~\citep{Cain2011}
\begin{equation}
\label{LensEqWithFlexion}
\boldsymbol{\hat{\beta}}_\textrm{c}=\boldsymbol{\theta}_\textrm{c}-\hat{\gamma}\boldsymbol{\theta}_\textrm{c}^*-\frac{1}{4}\hat{\mathcal{F}}^*\boldsymbol{\theta}_\textrm{c}^2-\frac{1}{2}\hat{\mathcal{F}}\theta^2-\frac{1}{4}\hat{\mathcal{G}}{\boldsymbol{\theta}_\textrm{c}^*}^2,
\end{equation}
where $^*$ denotes complex conjugation. Here $\gamma$, $\mathcal{F}$ and $\mathcal{G}$ are \textit{shear}, \textit{flexion} and \textit{second-flexion} variables, respectively (to be related to the lens potential later).

The shear coefficients, together with the convergence, correspond to the second-order derivatives of the lens potential. These terms are responsible for the magnification and elongation of an image relative to the source. If the gravitational field is strong enough for the shear to vary significantly over the cross-section of the image, higher-order effects do however come into play, creating a banana-looking shape. First in line are the so-called flexion terms, which correspond to a decomposition of the third-order derivative of the lensing potential, as we shall see below.

The effects of flexion are already discussed in~\cite{kaiser95a}, with the formalism properly developed in~\cite{goldberg2005}. A particularly tidy presentation in complex form is given in~\citet{Bacon06}.

In this paper we do not stop at flexion, but include even higher-order terms, as described next.

\subsection{Higher-Order Terms}
The Taylor expansion begun in the previous two sections could, of course, be extended to include even higher-order terms. A systematic route to including more and more higher-order terms is through the so-called \textit{null-geodesic deviation equation (GDE)}, which describes the deviation between two light rays whose separation is initially infinitesimal.
To first (second) order in the deviation vector $\boldsymbol{\nu}$, the null GDE provides the shear (flexion) terms. In 2016, however, Clarkson extended this analysis by keeping all leading-order screen-space derivatives~\citep{clarkson16a,clarkson16b}. The key point is the so-called roulette amplitudes of the lens, $\alpha_m^s(\psi)$ and $\beta_m^s(\psi)$, which encode the $(m+1)$-th order derivative of the lens potential. The roulette amplitudes in the weak-field, thin-lens paradigm are presented in real form in the aforementioned Appendix and in complex form below. Define the complex amplitude combination
\begin{equation}
\gamma_s^m=\alpha_s^m+i\beta_s^m.
\end{equation}
Also, define the complex derivative $\partial_\textrm{c}=\partial_1+i\partial_2$ and $\Box=\partial_\textrm{c}\partial_\textrm{c}^*$. A particularly useful expression for $\gamma_s^m$ is then given by
\begin{equation}
\label{GamNice1}
\gamma_s^m=\Gamma_s^m\Box^{a^{-}}\partial_c^s\psi,
\end{equation}
as shown in~\cite{normann20Rec}. Here $a^-=(m+1-s)/2$ and $\Gamma_s^m$ are numerical coefficients given by
\begin{equation}
\label{matrix}
\Gamma_s^m=
\begin{cases}
-\dfrac{1}{2^{m+\delta_{0s}}}\dbinom{m+1}{a^-}, & m+s\ \textrm{odd,}\\[1ex]
0, & \textrm{else.}
\end{cases}
\end{equation}
The number of non-zero amplitudes is shown in Figure~\ref{fig:ms-diagram}, for each $(m,s)$-tuple. To relate these to the previous discussion, we have the following relations~\citep{normann20Rec}:
\begin{align}
\label{relations}
\textrm{deflection}&=\left\{
\begin{aligned}
&\alpha^0_1=-\partial_1\psi\\
&\beta^0_1=-\partial_2\psi
\end{aligned}
\right.\\
\textrm{convergence $\&$ shear}&=\left\{
\begin{aligned}
  &\alpha^1_0=-\kappa=-\frac{1}{2}(\partial_{11}\psi+\partial_{22}\psi)\\
  &\alpha^1_2=-\gamma_+=-\frac{1}{2}(\partial_{11}\psi-\partial_{22}\psi)\\
&\beta^1_2=-\gamma_\times=-\partial_{21}\psi
\end{aligned}
\right.\\
\textrm{flexion terms}&=\left\{
\begin{aligned}
&\gamma^2_1=\mathcal{F}=-\frac{3}{4}\Box\partial_\textrm{c}\psi,\\
&\gamma^2_3=\mathcal{G}=-\frac{1}{4}\partial^3_\textrm{c}\psi,
\end{aligned}
\right.
\end{align}
Note that the roulette amplitude $\alpha_1^0$ (and $\beta_1^0$) above should not be confused with the reduced deflection angle $\boldsymbol{\alpha}=\nabla\psi$ of Eq.~(\ref{redDeflAng}): the two are closely related, but differ by a sign, $\alpha_1^0=-\partial_1\psi=-\alpha_1$. As shown in~\cite{normann20Rec}, there exist a number of recursion relations among these amplitudes, which have proven useful in the actual computation of these quantities for implementation in our simulator, as more thoroughly described elsewhere~\citep{cosmosim_wiki,hgs2023ecms}.

The roulette amplitudes presented above constitute our local expansion of the lens potential, and hence function as a way of quantifying the information content of a given image. For our purposes, we define:
\begin{Definition}
  The (useful) information content of an image is the number of non-zero roulette amplitudes that can be (usefully) extracted from it.
\end{Definition}
Hence, one way of increasing the information, is to increase the number of points around which one expands to low order. Another way is to increase the order of the expansion around a given point. The latter is the approach we take in this paper, and we will see that it is indeed possible to extract information beyond flexion terms.
The next step is to relate these amplitudes to the lens parameters, which is done in the following section.

\begin{figure}[H]
  \centering
\begin{tikzpicture}[
  x=1.0cm, y=0.85cm,
  dot/.style={font=\large\bfseries, text=green!70!black},
  lab/.style={font=\large}
]

% --- thick axes (the two lines) ---
\draw[line width=1pt] (0,0) -- (6.2,0);      % horizontal
\draw[line width=1pt] (0,0) -- (0,-7.2);     % vertical

% --- column labels m = 0..5 ---
\foreach \m in {0,1,2,3,4,5}{
  \node[lab] at (\m+0.5,0.45) {\m};
}

% --- row labels s = 0..6 ---
\foreach \s in {0,1,2,3,4,5,6}{
  \node[lab, anchor=east] at (-0.35,-\s-0.5) {\s};
}

% --- "m ->" and "s down" near the corner ---
\node[lab] at (-0.65,0.62) {$m$};
\draw[-{Stealth[length=3mm]}, line width=1.0pt] (-0.45,0.62) -- (-0.05,0.62);

\node[lab] at (-0.65,0.25) {$s$};
\draw[-{Stealth[length=3mm]}, line width=1.0pt] (-0.65,0.08) -- (-0.65,-0.18);

% --- numbers instead of dots ---

% s = 0: m = 1,3,5
\node[dot] at (1.5,-0.5) {1};
\node[dot] at (3.5,-0.5) {1};
\node[dot] at (5.5,-0.5) {1};

% s = 1: m = 0,2,4
\node[dot] at (0.5,-1.5) {2};
\node[dot] at (2.5,-1.5) {2};
\node[dot] at (4.5,-1.5) {2};

% s = 2: m = 1,3,5
\node[dot] at (1.5,-2.5) {2};
\node[dot] at (3.5,-2.5) {2};
\node[dot] at (5.5,-2.5) {2};

% s = 3: m = 2,4
\node[dot] at (2.5,-3.5) {2};
\node[dot] at (4.5,-3.5) {2};

% s = 4: m = 3,5
\node[dot] at (3.5,-4.5) {2};
\node[dot] at (5.5,-4.5) {2};

% s = 5: m = 4
\node[dot] at (4.5,-5.5) {2};

% s = 6: m = 5
\node[dot] at (5.5,-6.5) {2};

\end{tikzpicture}
\caption{The green numbers indicate which $(m,s)$ combinations are allowed, and how many parameters there are.\label{fig:ms-diagram}}
\end{figure}
In order to devise the most interesting experimental setups, it will be useful to know how the roulette amplitudes evolve. For the symmetric case, it has been shown~\cite{clarkson16b} that the roulette amplitudes fall off as
$\xi^{m-2}$ for the SIS lens, which is one order of $\xi$ slower than for the point-mass lens. We will therefore choose the SIS lens and its elliptical extension, the SIE lens, as the basis for our test scenarios. These lenses are described in more detail in the following section.

\section{Materials and Methods}
\label{sec:method}
\begin{figure}[H]
    \centering
    \resizebox{\linewidth}{!}{%
    \begin{tikzpicture}[
      >=Latex,
      every node/.style={font=\small},
      box/.style={
        rectangle, draw=black, thick, minimum width=3.4cm, minimum height=1.1cm,
        align=center, fill=boxgray
      },
      circ/.style={
        circle, draw=black, thick, minimum size=1.3cm, align=center
      },
      arr/.style={-{Latex[length=3mm]}, line width=1pt},
      arrL/.style={-{Latex[length=3mm]}, line width=1pt, rounded corners=4pt},
      noisebox/.style={
        rectangle, draw=black, thick, align=center, fill=noisepurple,
        inner sep=6pt
      }
    ]
    % ---- Absolute positions (cm) ----
    \node[circ, fill=boxgray]                               (lamGT)    at (3.4,4.2) {$\boldsymbol{\lambda}_\text{GT}$};
    \node[box, fill=cosmoblue, text=cosmobluetext]          (ampcalc)  at (3.4,2.6) {AMPLITUDE\\CALCULATION};
    \node[circ, fill=boxgray]                               (gammaGT)  at (3.4,1.1) {$\gamma_\text{GT}$};
    \node[box, fill=cosmoblue, text=cosmobluetext, minimum width=1.9cm, minimum height=0.9cm] (noise)    at (1.7,-0.05){NOISE};
    \node[circ, fill=boxgray]                               (gammai)   at (3.4,-1.2){$\gamma^*$};
    \node[box, fill=cosmoblue, text=cosmobluetext]          (opt)      at (7.4,-1.2){OPTIMIZATION\\PROCEDURE};
    \node[circ, fill=boxgray]                               (lamstar)  at (7.4,1.1){$\boldsymbol{\lambda}^{\ast}$};
    \node[box, fill=cosmoblue, text=cosmobluetext, minimum width=1.4cm] (rsq)      at (7.4,4.2) {$R^2$};
    \coordinate (entryPt) at (1.2,4.2);
    \coordinate (exitPt)  at (9.4,4.2);
    % ---- Edges ----
    \coordinate (gimid) at ($(gammaGT)!0.5!(gammai)$);
    \draw[arr, dashed] (entryPt) -- (lamGT);
    \draw[arr] (lamGT) -- (ampcalc);
    \draw[arr] (ampcalc) -- (gammaGT);
    \draw[arr] (lamGT) -- (rsq);
    \draw[arr] (gammaGT) -- (gammai);
    \draw[arr] (noise.east) -- (gimid);
    \draw[arr] (gammai) -- (opt);
    \draw[arr] (opt) -- (lamstar);
    \draw[arr] (lamstar) -- (rsq);
    \draw[arr, dashed] (rsq) -- (exitPt);
    \end{tikzpicture}%
    }
    \caption{Process diagram of the pipeline, from leftmost (dashed) arrow to rightmost (dashed) arrow. The ground-truth lens parameters $\boldsymbol{\lambda}_\text{GT}$ are used to calculate the corresponding roulette amplitudes $\gamma_\text{GT}$. Adding noise yields the observed amplitudes $\gamma^*$, which enter the optimization procedure and return the recovered parameters $\boldsymbol{\lambda}^{\ast}$. The recovery is validated by comparing $\boldsymbol{\lambda}^{\ast}$ to the true $\boldsymbol{\lambda}_\text{GT}$ using $R^2$ with bootstrapping.\label{fig:floatdiagr}}
\end{figure}
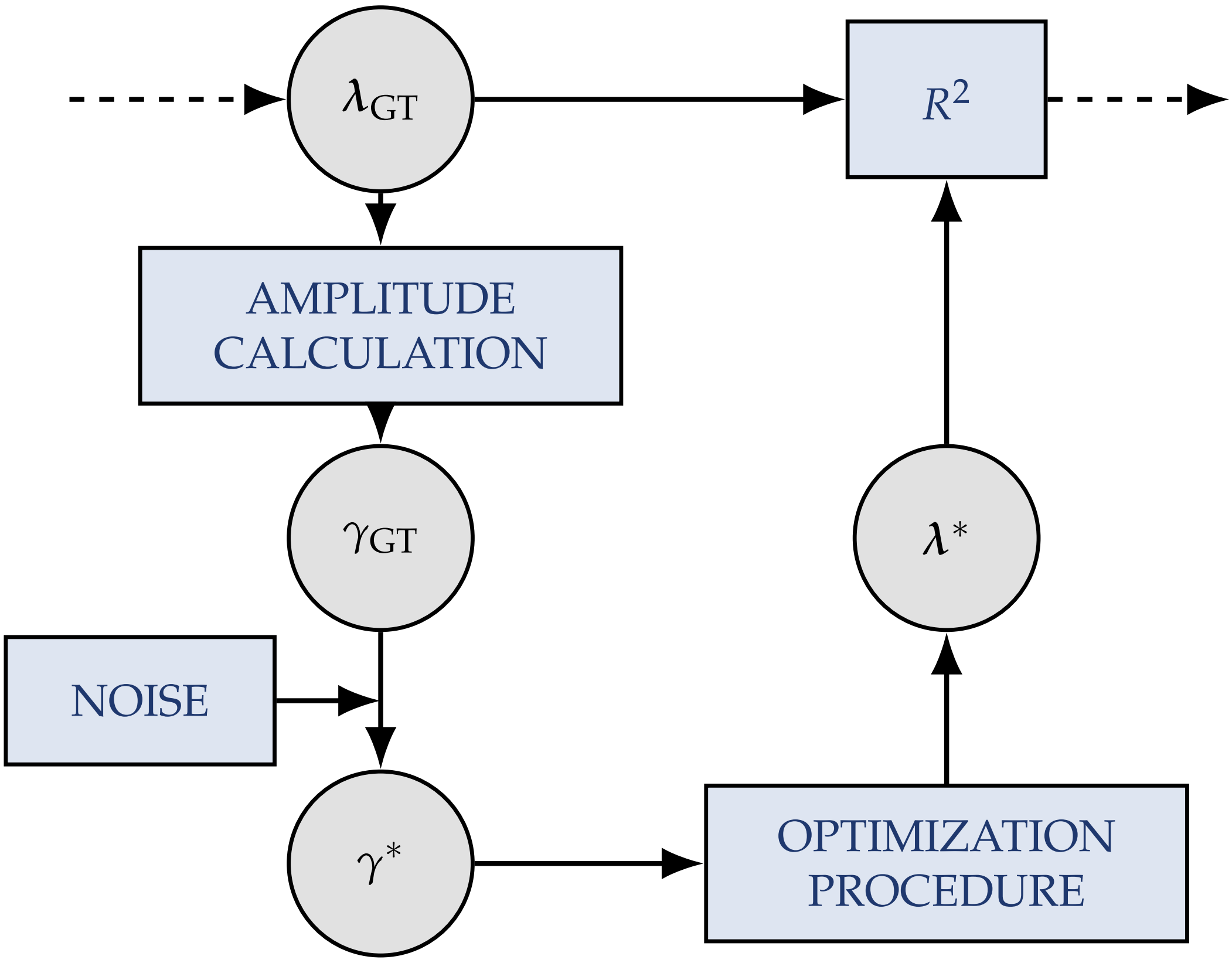
In the present work we investigate the ability to perform lens-mass reconstruction from the data of one single image. The amount of information needed in order to reconstruct the lens mass is determined by the complexity of the lensing scenario. The key question, as posed in our research question, is therefore how much data is available in an image. To find out, we have devised the process model presented in Figure~\ref{fig:floatdiagr}. Steps A and B have already been established in prior work and form the foundation for the present paper; the present paper's contribution lies in Steps C and D, described below.
\begin{itemize}
\item \textcolor{green!60!black}{\ding{51}}\textit{Step A (simulator)}: CosmoSim, our simulator, provides an image-to-amplitude capability: given a lens model $\psi_{\boldsymbol{\lambda}}$, it returns the corresponding roulette amplitudes $\gamma_\text{GT}$ (GT for Ground Truth). This gives an in-principle ability to retrieve data beyond flexion in the absence of noise. Further details may be found in~\citep{cosmosim_wiki,hgs2023ecms}.
\item \textcolor{green!60!black}{\ding{51}}\textit{Step B (real data)}: For the simulator to be of any practical use, it should be able to reconstruct real data. This is the case, as reported in~\citep{oda2026ngc6505}.
\item \textcolor{steparrow}{\ding{220}}\textit{Step C (Add noise)}: Real amplitude measurements are never noise-free. In the process we envisage, the dominant noise source is measurement/extraction error -- e.g., the prediction error of a machine-learning-based amplitude extractor (cf.~\citealt{hgs2023nik}). We simulate this by adding synthetic noise to $\gamma_\text{GT}$, allowing us to assess how reconstruction performance degrades as a function of noise level.
\item \textcolor{steparrow}{\ding{220}}\textit{Step D (Lens-mass reconstruction)}: In this step we investigate at what order of $m$ the lens-mass reconstruction ceases to improve.
\end{itemize}

In this work, we concentrate on Steps C and D, and investigate the ability to do lens-mass reconstruction from noisy data. Concretely, we investigate whether or not higher-order terms are beneficial for lens-mass reconstruction, and if so, under what circumstances. We will next describe the scenarios investigated.

\subsection{Scenarios Investigated} \label{scenarios_section}
In order to answer our research question we will investigate both single-lens scenarios, and also cluster lenses. We investigate the following different scenarios:

\paragraph{\textbf{Scenario 1}} \textit{Ability to do lens-mass reconstruction in the case of a single Singular Isothermal Ellipsoid (SIE) lens;} This lens is characterized by five parameters: the Einstein radius $\theta_\text{E}$, the two position coordinates of the lens $(\theta_1, \theta_2)$, the orientation $\phi_0$ and the ellipticity $f$ of the lens. The lens potential is given by
    \begin{align}
    \label{psiSIS}\psi_\textrm{SIE}=\theta_\text{E}\frac{\sqrt{f}}{f^{\prime}}\left(\theta_2^\prime\arcsin\left(f^\prime\frac{\theta_2^\prime}{\theta}\right) +\theta_1^\prime\textrm{arsinh}\left(\frac{f^{\prime}}{f}\frac{\theta_1^\prime}{\theta}\right)\right)
    \end{align}
    where $\theta=\sqrt{\theta_1^2+\theta_2^2}$ is the radial distance from the center of the lens and $(\theta_1^\prime,\theta_2^\prime)$ are the coordinates of the lens rotated through an angle $\phi_0$. Moreover, $f^\prime=\sqrt{1-f^2}$ is the complementary ellipticity.
    The hypothesis is that reconstruction from shear and flexion data alone (i.e.\ up to third order) will perform just as well as reconstruction that also includes higher-order terms, since there are only five independent parameters to recover. As such, this scenario serves as a sanity check for our methodology.

   \paragraph{\textbf{Scenario 2}}\textit{Ability to do lens-mass reconstruction in the case of multiple SIE lenses.} In this case the lens potential is given by
\begin{align}
    \label{psiSIEcluster}\psi_\textrm{nSIE}=\sum_{i=1}^n\psi_{\textrm{SIE},i}
    \end{align}
    where each SIE component $\psi_{\textrm{SIE},i}$ is characterized by 5 parameters. Thus the total number of parameters is $N_p=5\times n$. From Figure~\ref{fig:ms-diagram} it is clear that there are 9 roulette amplitudes up to and including flexion ($m=2$). From the formula for $N_p$, this suffices for a single lens, but not quite for a two-lens system and beyond ($n\geq 2$). Our hypothesis is therefore that higher-order terms will aid the analysis in the case of multiple lenses. We will test $n=2$. To assess the impact of noise on this hypothesis, this scenario is tested both without noise (\textbf{Scenario 2a}) and with noise (\textbf{Scenario 2b}); in the latter case, we hypothesize that the reconstructed parameters will deviate more from their true value than in Scenario 2a, and that this deviation will be reduced when higher-order terms are included.

    \paragraph{\textbf{Scenario 3}}\textit{Ability to do lens-mass reconstruction in the case of multiple SIS lenses;}
    \begin{align}
    \label{psiSIScluster}\psi_\textrm{SIS}=\theta_\text{E}\theta\quad\text{and}\quad\psi_\textrm{nSIS}=\sum_{i=1}^n\psi_{\textrm{SIS}i}.
    \end{align}
    where $\theta=\sqrt{\theta_1^2+\theta_2^2}$ is the radial distance from the center of the lens. In this case the number of parameters is $N_p=3\times n$. By the same type of argument as for the SIE cluster, we will in this case need terms beyond flexion whenever $n\geq 4$. We choose $n=4$ and hypothesize that higher-order terms will improve the lens-mass reconstruction. As for Scenario 2, this scenario is tested both without noise (\textbf{Scenario 3a}) and with noise (\textbf{Scenario 3b}).

\subsection{Synthesized Dataset}
\label{cluster_lens_method}

The lenses used are randomly generated. The Einstein radii for our lenses are between $5$ and $100$ units. Their position is between $-100$ and $100$ units along both axes. The rotation is between $0^\circ$ and $180^\circ$, since the SIE model is periodic with a period of $180^\circ$. Finally, the ellipse ratio $f$ is restricted to the interval $0.1\,\leq f \leq 0.9$, so as to accommodate very elongated lenses as well as near-spherical ones. The position of the lenses is also kept more than 30 units away from the point at which the roulette amplitudes are calculated ($\boldsymbol{\xi}$) to avoid some lenses making the impact of others insignificant due to the relatively high value of their roulette amplitudes. In the case of the SIS model, only the Einstein radius and position are needed, so no rotation or ellipse ratio is generated. The ranges of the values used are largely chosen to align with Cosmosim.

\paragraph{Noise modeling} We apply a layer of Gaussian noise to the roulette amplitudes. The noisy amplitudes have a mean equal to the true amplitude and a standard deviation equal to a percentage of the absolute value of the true amplitude; for the single SIE lens, this percentage is $15\%$ for the first-, second-, and third-order amplitudes, $20\%$ for the fourth-order amplitudes, and $25\%$ for the fifth-order amplitudes. For the cluster lenses, we use a third of these values. It is important to note that the lenses, and the noisy roulette amplitudes derived from them, are the same across all tests conducted within a scenario. All roulette amplitudes are calculated initially, with the higher orders simply excluded from consideration in the initial tests within a scenario.

\paragraph{Cluster lenses}

The reader should note that cluster lenses are easily implemented from the one-lens data due to the linearity of the roulette amplitudes. For a cluster lens $\psi=\sum_i^n\psi_i$, we have roulette amplitudes given by
\begin{align*}
\gamma(\psi)=\gamma(\psi_1)+...+\gamma(\psi_n)
\end{align*}
This allows for separate generation of effects from each member of the cluster.

\subsection{Lens-Mass Reconstruction}

Given a set of observed, possibly noisy, roulette amplitudes $\gamma^*$, the goal of this subsection is to recover the underlying lens parameters $\boldsymbol{\lambda}$ via a regression analysis: we minimize a least-squares sum between the observed amplitudes and those predicted by a candidate lens model, computed at a single point in the image. From the general roulette-amplitude formula, Eq.~\eqref{GamNice1}, we calculate the predicted roulette amplitudes for a particular type of lens.

We minimize a least-squares sum to acquire our result, as is typical of a problem of this form. Note, however, that this procedure is applied to each roulette amplitude individually, since they take different functional forms. In each case the lens parameters $\boldsymbol{\lambda}$ are the parameters of the regression. Moreover, as the absolute value of the roulette amplitudes ($\gamma^m_s$) typically decrease by an order of magnitude for each order of $m$, a weighting term is added so that all roulette amplitudes affect the final sum roughly equally. Finally, to keep notation at a minimum, we henceforth stack all the roulette amplitudes into a single array, $(\alpha^0_1,\beta^0_1,\alpha^1_0,\beta^1_0,\alpha^1_2,\beta^1_2,...)$, and refer to the $i$th component of this array as $\gamma_i$. The least-squares sum $S$ is therefore
\begin{equation}
    S=\sum^N_{i=1}\frac{1}{\overline{\gamma}_i^2}(\gamma_i^*-\gamma_i(\boldsymbol{\lambda}))^2
\end{equation}
where
\begin{itemize}
    \item $\gamma_ i^*$ is input roulette amplitude $i$ (with noise).
    \item $\gamma_i(\boldsymbol{\lambda})$ is the model's predicted value for the roulette amplitude of the same index as $\gamma_i^*$.
    \item $\boldsymbol{\lambda}$ are the lens parameters.
    \item $\overline{\gamma}_i$ is the arithmetic mean of the roulette amplitudes of the same order as $\gamma_i^*$, and serves as the basis of the weighting term.
\end{itemize}

In summary, by minimizing $S$ we can get a set of lens parameters from a set of roulette amplitudes. To minimize $S$ with respect to the lens parameters, we have developed a Python script that uses the \texttt{minimize} function from the \texttt{scipy.optimize} library~\citep{Virtanen2020}. We specifically use the Trust-Region Constrained Algorithm. To limit the range of possible parameters to reasonable constraints, bounds are implemented. The Einstein radius is capped between $1$ and $150$ units, and the ellipse ratio $f$ is capped between $0.05$ and $0.95$. Rotation is unbound, but wrapped to the range $0$ to $180^\circ$, since that is the SIE model's period.

Regarding the position, we assume that lenses contain some visible matter, meaning that an approximate position could be provided in a real scenario. To approximate this, a point is chosen randomly within a radius of $25$ units, with bounds of $\pm 30$ units along each axis centered on this point. 3 different starting points for the regression analysis are chosen beforehand within this region. The best result is then chosen as the final result, determined by which starting point produces roulette amplitudes most similar to the initial set of roulette amplitudes. In the case of multiple lenses, each combination of these starting points is tried. Table~\ref{tab:summary_of_bounds} summarizes the resulting ranges of the generated lens parameters, the bounds used in the minimization, and the starting points.

The number of starting points was determined through trial and error while testing on a single SIE lens, with the aim of consistently finding the correct local minimum with the fewest possible points to reduce runtime. The regression analysis is still not wholly reliable, occasionally finding the wrong local minimum even without noise; this is more pronounced for cluster lenses.

Finally, in the case of cluster lenses, there is no guarantee that the resulting lenses appear in the same order as the true lenses. For instance we may have a two lens cluster with Lens 1 and Lens 2. The order in which their roulette amplitudes are added to the final set of roulette amplitudes won't affect the final sum. Therefore, we may get the estimated lens parameters in either the order Lens 1 - Lens 2 or Lens 2 - Lens 1. When the true and estimated lens parameters are then compared, we might end up comparing Lens 1 to Lens 2 (and vice versa) and get wrong results. To avoid this, we check to see which configuration results in the smallest total distance between the true and the estimated lens parameters.

\begin{table}[H]
    \centering
    \caption{Summary of the ranges of the lens parameters of the generated lenses and the minimization bounds, as well as the starting points. The rotation and ellipse ratio $f$ does not matter for the SIS-model.\label{tab:summary_of_bounds}}
    \begin{tabular}{lccc}
        \toprule
         & \textbf{Generated Lenses} & \textbf{Bounds for Minimization} & \textbf{Starting Points} \\
        \midrule
        $\theta_\textrm{E}$ & $[10,100]$ & $[1,150]$ & $50$ or $5$ \\
        Position & $[-100,100]$ & varies & varies \\
        Rotation & $[0^\circ,180^\circ]$ & unbounded & $90^\circ$ \\
        $f$ & $[0.1,0.9]$ & $[0.05,0.95]$ & $0.5$ \\
        \bottomrule
    \end{tabular}
\end{table}

\subsection{Evaluation Metric} \label{sec:metric}

To quantify the accuracy of the lens-mass reconstruction, we use the coefficient of determination $R^2$, with the ground truth as the reference model. For each lens parameter $\lambda\in\boldsymbol{\lambda}$ (e.g.\ the Einstein radius $\theta_\textrm{E}$, the position, the rotation, or the ellipticity $f$) we evaluate
\begin{equation}
R^2=1-\frac{SS_\textrm{res}}{SS_\textrm{tot}}
\end{equation}
where $SS_\textrm{res}$ is the sum of squares of residuals and $SS_\textrm{tot}$ is the total sum of squares, defined by
\begin{eqnarray}
    & SS_\textrm{tot}=\sum_i(\lambda^*_i-\overline{\lambda^*})^2\nonumber\\
    & SS_\textrm{res}=\sum_i(\lambda^*_i-\lambda_{\text{GT},i})^2\nonumber\\
\end{eqnarray}
Here the index $i=1,\dots,N$ runs over the lenses in one bootstrap sample (see below); $\lambda_i^*$ is the recovered value of the parameter $\lambda$ for lens $i$, $\lambda_{\text{GT},i}$ is its corresponding ground-truth value, and $\overline{\lambda^*}$ is the arithmetic mean of $\lambda_i^*$ over the sample.

We use bootstrapping to calculate the coefficient of determination and its uncertainty. For each test, 500 lenses are generated and their parameters estimated (see Sect.~\ref{sec:results}); from these 500 estimated lens parameters, we create 1000 sets of 100 lenses each, chosen randomly (with replacement). For each set, we discard the 5\% best and worst results and calculate $R^2$; the values reported throughout Sect.~\ref{sec:results} are the mean and standard deviation of $R^2$ over the remaining 900 sets.

\section{Results}
\label{sec:results}
We evaluate the accuracy of the lens-mass reconstruction using the coefficient of determination $R^2$, computed via bootstrapping as described in Sect.~\ref{sec:metric}.

For each scenario studied below, we compare three different tests (summarized, together with the five scenarios, in Table~\ref{tab:scenario_test_overview}). Throughout this paper, \textit{order} refers to the derivative order of the lensing potential $\psi$, i.e.\ $m+1$ in the roulette-amplitude notation of Sect.~\ref{sec:theory}, and not the roulette index $m$ itself; third order thus coincides with flexion (and $m=2$).
\begin{itemize}[itemsep=1pt, topsep=2pt, parsep=0pt]
    \item \textit{Test 1}: Using amplitudes up to and including the third order.
    \item \textit{Test 2}: Using amplitudes up to and including the fourth order.
    \item \textit{Test 3}: Using amplitudes up to and including the fifth order.
\end{itemize}

\begin{table}[H]
    \centering
    \caption{Overview of the five scenarios (Sect.~\ref{sec:method}) and the three tests (amplitudes truncated at 3rd, 4th, or 5th order) applied to each. Scenarios 2 and 3 (SIE and SIS cluster lenses) are each tested both without noise (2a, 3a) and with noise (2b, 3b); all three tests are run for every row. The Scenario column also lists the table with the corresponding $R^2$-values and the appendix with the corresponding scatter plots.\label{tab:scenario_test_overview}}
    \begin{adjustwidth}{-\extralength}{0cm}
    \centering
    \begin{tabularx}{\fulllength}{CCCCCC}
        \toprule
        \textbf{Scenario [Text ref.]} & \textbf{Lens (nbr. param.)} & \textbf{Noise} & \textbf{Test 1} & \textbf{Test 2} & \textbf{Test 3} \\
        & & & ($\leq$3rd order) & ($\leq$4th order) & ($\leq$5th order) \\
        \midrule
        Scen.~1:~Table~\ref{tab:scenario1_results} and App.~\ref{app:scenario1_plots} & 1$\times$SIE (5~param.) & y & y & y & y \\
        \noalign{\vskip 2pt\color{gray}\hrule height 0.4pt\vskip 2pt}
        Scen.~2a:~Table~\ref{tab:scenario2a_results} and App.~\ref{app:scenario2_plots} & 2$\times$SIE (10~param.) & \textcolor{gray}{n} & y & y & y \\
        \noalign{\vskip 2pt\color{gray}\hrule height 0.4pt\vskip 2pt}
        Scen.~2b:~Table~\ref{tab:scenario2b_results} and App.~\ref{app:scenario2_plots} & 2$\times$SIE (10~param.) & y & y & y & y \\
        \noalign{\vskip 2pt\color{gray}\hrule height 0.4pt\vskip 2pt}
        Scen.~3a:~Table~\ref{tab:scenario3a_results} and App.~\ref{app:scenario3_plots} & 4$\times$SIS (12~param.) & \textcolor{gray}{n} & y & y & y \\
        \noalign{\vskip 2pt\color{gray}\hrule height 0.4pt\vskip 2pt}
        Scen.~3b:~Table~\ref{tab:scenario3b_results} and App.~\ref{app:scenario3_plots} & 4$\times$SIS (12~param.) & y & y & y & y \\
        \bottomrule
    \end{tabularx}
    \end{adjustwidth}
\end{table}

\subsection{Analysis of Scenario 1: Single SIE Lens}

The results are summarized in Table~\ref{tab:scenario1_results}. The corresponding scatter plots for all three tests are presented in Appendix~\ref{app:scenario1_plots}.

\begin{figure}[H]
    \centering
    \begin{subfigure}{0.4\linewidth}
        \centering
        \includegraphics[width=\linewidth]{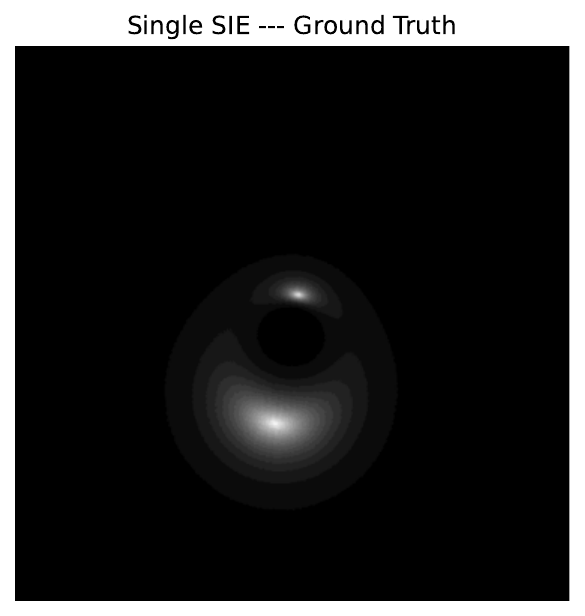}
    \end{subfigure}
    \begin{subfigure}{0.4\linewidth}
        \centering
        \includegraphics[width=\linewidth]{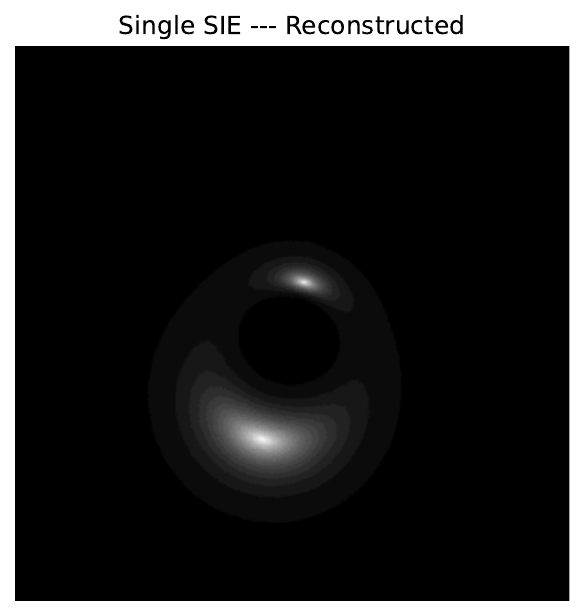}
    \end{subfigure}
    \caption{This is a visual example of Test 2 of Scenario 1. On the left is the result of using the true parameters, on the right is the result of using the estimated parameters after noise is applied. Both images share the same spherical light source with a S\'ersic profile and are generated using ray-tracing.\label{fig:lens_comp}}
\end{figure}

Recall that, for this five-parameter lens, we hypothesized (Sect.~\ref{scenarios_section}) that third-order (shear-and-flexion) amplitudes alone would suffice, and that higher-order terms would therefore not noticeably improve the reconstruction. Contrary to this hypothesis, we observe a notable improvement in the accuracy of the parameters (in particular the rotation angle and the ellipse ratio) when using fourth- and fifth-order amplitudes, compared to using only up to and including third-order amplitudes. Beyond this, we can see from the scatter plots (Figure~\ref{fig:app_scenario1_grid}) that the rotation estimate for some lenses with an ellipse ratio close to 1 skews heavily away from the ground truth, which is the cause of the extreme inaccuracy in the $R^2$-value for the rotation. This is likely because of those lenses being near spherical, so the rotation has a much lower impact. Figure~\ref{fig:lens_comp} shows a visual example of the lens recovery for Test 2 of this scenario.

\begin{table}[H]
    \centering
    \caption{$R^2$-values (mean $\pm$ bootstrap standard deviation) for Scenario 1 (single SIE lens), for the three tests defined in Table~\ref{tab:scenario_test_overview}. The scatter plots underlying these values are shown in Appendix~\ref{app:scenario1_plots}.\label{tab:scenario1_results}}
    \begin{tabular}{lccc}
        \toprule
        \textbf{Parameter} & \textbf{1} ($\leq$3rd) & \textbf{2} ($\leq$4th) & \textbf{3} ($\leq$5th) \\
        \midrule
        $\theta_\textrm{E}$ (\%) & $89.86\pm1.83$ & $90.73\pm1.94$ & $88.86\pm2.66$ \\
        Position (\%) & $97.61\pm0.24$ & $98.83\pm0.18$ & $99.19\pm0.16$ \\
        sin(Rot.) (\%) & $39.28\pm13.50$ & $47.37\pm12.93$ & $68.98\pm11.42$ \\
        $f$ (\%) & $52.84\pm7.69$ & $72.05\pm6.45$ & $79.82\pm6.02$ \\
        \bottomrule
    \end{tabular}
\end{table}

\subsection{Analysis of Scenario 2: SIE Cluster Lenses}

There are 9 roulette amplitudes up to and including third-order terms. As a necessary-condition heuristic, we treat each roulette amplitude as providing one independent constraint, so that the most complex lens fittable from this many amplitudes has at most 9 independent parameters; this parameter-counting argument does not by itself rule out degeneracies for specific lens configurations, which we instead assess empirically below. A SIE-model lens has 5 parameters in total, meaning that a single SIE-model lens is at the boundary of what can be derived from this many roulette amplitudes. If we additionally use fourth-order roulette amplitudes, we have 14 in total, in which case a 2-SIE-model lens cluster could in theory be derived from one set of roulette amplitudes alone. We therefore try a test similar to before, but with the lenses clustered into pairs. The parameters of the roulette amplitudes are chosen randomly, with the true roulette amplitudes calculated and summed together to create the set for the cluster lens, as shown in Sect.~\ref{cluster_lens_method}. We use the same 500 lenses for each test within a scenario, though the tests with and without noise use different sets of lenses. The results are summarized in Tables~\ref{tab:scenario2a_results} and~\ref{tab:scenario2b_results}. The corresponding scatter plots for all three tests, with and without noise, are presented in Appendix~\ref{app:scenario2_plots}.

Without noise (Scenario 2a), the Einstein radius and position are already recovered reasonably well at third order, and accuracy improves further with higher-order terms. The rotation and ellipse ratio, meanwhile, are recovered poorly at third order, with the accuracy improving starkly when higher-order amplitudes are used. As can be seen in the corresponding scatter plot, most lenses perfectly correspond to their true value in Test 2 and Test 3, but a number of lenses which found the wrong local minimum still cause a significant impact on the $R^2$-value.

With noise added (Scenario 2b), using only up to and including third-order roulette amplitudes gives very inaccurate results; it is important to note that the position is heavily constrained, so it will always be accurate to a certain degree. Including fourth-order amplitudes massively increases the accuracy of the parameters, and including fifth-order amplitudes gives a generally smaller further increase across most parameters. As in Scenario 2a, the impact of additional higher-order roulette amplitudes is most notable for the rotation and the ellipse ratio. Comparing the two directly, noise substantially degrades the reconstruction accuracy for every parameter and test, most severely for the rotation and ellipse ratio at third order, where the noisy case even yields negative $R^2$-values. For both these scenarios, the issues with the rotation identified in Scenario 1 remain.

\begin{table}[H]
    \centering
    \caption{$R^2$-values (mean $\pm$ bootstrap standard deviation) for Scenario 2a (SIE cluster, $n=2$, no noise), for the three tests defined in Table~\ref{tab:scenario_test_overview}. The scatter plots underlying these values are shown in Appendix~\ref{app:scenario2_plots}.\label{tab:scenario2a_results}}
    \begin{tabular}{lccc}
        \toprule
        \textbf{Parameter} & \textbf{1} ($\leq$3rd) & \textbf{2} ($\leq$4th) & \textbf{3} ($\leq$5th) \\
        \midrule
        $\theta_\textrm{E}$ (\%) & $70.79\pm4.86$ & $89.34\pm3.39$ & $93.71\pm2.70$ \\
        Position (\%) & $98.15\pm0.22$ & $99.27\pm0.17$ & $99.40\pm0.20$ \\
        sin(Rot.) (\%) & $20.68\pm9.38$ & $79.51\pm5.66$ & $87.16\pm4.61$ \\
        $f$ (\%) & $21.43\pm9.29$ & $78.39\pm5.94$ & $85.60\pm4.76$ \\
        \bottomrule
    \end{tabular}
\end{table}

\begin{table}[H]
    \centering
    \caption{$R^2$-values (mean $\pm$ bootstrap standard deviation) for Scenario 2b (SIE cluster, $n=2$, with noise), for the three tests defined in Table~\ref{tab:scenario_test_overview}. The scatter plots underlying these values are shown in Appendix~\ref{app:scenario2_plots}.\label{tab:scenario2b_results}}
    \begin{tabular}{lccc}
        \toprule
        \textbf{Parameter} & \textbf{1} ($\leq$3rd) & \textbf{2} ($\leq$4th) & \textbf{3} ($\leq$5th) \\
        \midrule
        $\theta_\textrm{E}$ (\%) & $44.56\pm6.19$ & $61.50\pm6.42$ & $72.79\pm5.93$ \\
        Position (\%) & $93.61\pm0.44$ & $95.56\pm0.40$ & $96.81\pm0.35$ \\
        sin(Rot.) (\%) & $-10.70\pm10.61$ & $27.99\pm9.69$ & $39.99\pm8.71$ \\
        $f$ (\%) & $-26.95\pm11.68$ & $45.66\pm6.71$ & $61.16\pm5.68$ \\
        \bottomrule
    \end{tabular}
\end{table}

\subsection{Analysis of Scenario 3: SIS Cluster Lenses}

Four SIS-model lenses have a total of 12 parameters. As stated in the previous tests, there are 9 roulette amplitudes up to and including the third order, which goes up to 14 when fourth-order roulette amplitudes are included. The test is conducted the same way as for the previous cluster lens, but with 4 SIS-model lenses instead. We use the same 500 lenses for each test within a scenario, though the tests with and without noise use different sets of lenses. The results are summarized in Tables~\ref{tab:scenario3a_results} and~\ref{tab:scenario3b_results}. The corresponding scatter plots for all three tests, with and without noise, are presented in Appendix~\ref{app:scenario3_plots}.

Without noise (Scenario 3a), the accuracy of the position is already very high at third order and stays roughly constant across tests, while the accuracy of the Einstein radius improves markedly when adding more roulette amplitudes. Similar to Scenario 2a, most lenses are at (or close to) ground truth in Test 2 and Test 3, but some lenses which failed to find the correct local minimum still cause a significant impact on the $R^2$-value.

With noise added (Scenario 3b), the same qualitative pattern holds --- the accuracy of the position does not change much between the tests, while the Einstein radius improves with higher-order terms --- but at a substantially lower overall accuracy. Comparing the two directly, noise has a comparatively small effect on the position (which remains above $91\%$ in both cases) but a large effect on the Einstein radius, whose $R^2$-value is roughly halved or worse at every test order once noise is added.

\begin{table}[H]
    \centering
    \caption{$R^2$-values (mean $\pm$ bootstrap standard deviation) for Scenario 3a (SIS cluster, $n=4$, no noise), for the three tests defined in Table~\ref{tab:scenario_test_overview}. The scatter plots underlying these values are shown in Appendix~\ref{app:scenario3_plots}.\label{tab:scenario3a_results}}
    \begin{tabular}{lccc}
        \toprule
        \textbf{Parameter} & \textbf{1} ($\leq$3rd) & \textbf{2} ($\leq$4th) & \textbf{3} ($\leq$5th) \\
        \midrule
        $\theta_\textrm{E}$ (\%) & $60.46\pm3.74$ & $94.53\pm1.11$ & $96.90\pm1.08$ \\
        Position (\%) & $97.79\pm0.14$ & $99.88\pm0.03$ & $99.96\pm0.05$ \\
        \bottomrule
    \end{tabular}
\end{table}

\begin{table}[H]
    \centering
    \caption{$R^2$-values (mean $\pm$ bootstrap standard deviation) for Scenario 3b (SIS cluster, $n=4$, with noise), for the three tests defined in Table~\ref{tab:scenario_test_overview}. The scatter plots underlying these values are shown in Appendix~\ref{app:scenario3_plots}.\label{tab:scenario3b_results}}
    \begin{tabular}{lccc}
        \toprule
        \textbf{Parameter} & \textbf{1} ($\leq$3rd) & \textbf{2} ($\leq$4th) & \textbf{3} ($\leq$5th) \\
        \midrule
        $\theta_\textrm{E}$ (\%) & $27.74\pm4.57$ & $40.68\pm3.79$ & $41.09\pm4.29$ \\
        Position (\%) & $91.21\pm0.33$ & $91.57\pm0.32$ & $91.60\pm0.35$ \\
        \bottomrule
    \end{tabular}
\end{table}

\section{Conclusions}
\label{sec:conclusion}

In this work we have investigated the effect of weak-lensing data beyond flexion for lens-mass reconstruction in the context of parametric models. In order to answer our research question we investigated the ability to perform lens-mass reconstruction from roulette amplitudes truncated at third (flexion), fourth, and fifth order, across five scenarios: a single SIE lens (Scenario 1), SIE cluster lenses with and without noise (Scenarios 2a--2b), and SIS cluster lenses with and without noise (Scenarios 3a--3b).

Across all scenarios, we find that higher-order amplitudes do provide additional, useful information for lens-mass reconstruction, and that this benefit is most pronounced exactly where it matters: for cluster lenses, and in the presence of noise. For the single SIE lens (Scenario 1), where the five parameters are, in principle, already recoverable from third-order data, including fourth- and fifth-order amplitudes still measurably improves the rotation and ellipticity estimates. For SIE cluster lenses (Scenario 2), the improvement is substantial: without noise, the rotation and ellipticity are poorly recovered at third order but improve starkly at fourth and fifth order; with noise added, third-order reconstruction becomes unreliable (even yielding negative $R^2$-values for rotation and ellipticity), while higher-order amplitudes restore much of the lost accuracy. For SIS cluster lenses (Scenario 3), the Einstein radius shows a similar pattern, improving markedly with higher-order amplitudes both with and without noise, while the (more tightly constrained) position is comparatively insensitive to the amplitude order. In short, the more complex the lens system and the noisier the data, the more beneficial it is to draw data beyond flexion.

These results come with several limitations. First, we have deliberately restricted ourselves to synthetic data, since our research question concerns whether higher-order terms are useful \textit{in principle}; extending this to real observational data, building on our related work on CosmoSim~\citep{oda2026ngc6505}, is a natural and necessary next step. Second, our parameter-recovery procedure is a least-squares regression with a small number of starting points, and is not wholly reliable even without noise; a more robust optimization strategy, or an analytic treatment of parameter degeneracies, would strengthen the conclusions drawn here. Third, our noise model is a simplified Gaussian approximation; as discussed in Sect.~\ref{sec:method}, a more realistic noise source in the proposed pipeline would stem from measurement or extraction error, such as the prediction error of a machine-learning-based amplitude extractor, and modeling that explicitly is left for future work.

Looking ahead, a more comprehensive approach would involve a detailed (analytic) analysis of higher-order terms for cluster lenses, to help identify the most promising observational setups to target. Combined with an extension to real observational data, this would allow for a more robust assessment of the practical utility of higher-order lensing information in the study of cluster mass distributions.

%=================================================================
% Back matter
\authorcontributions{Conceptualization, B.D.N. and H.G.S.; methodology, B.D.N. and H.G.S.; software, M.H.E.; formal analysis, M.H.E.; investigation, M.H.E.; resources, N.O.P.V.; data curation, N.O.P.V.; writing---original draft preparation, B.D.N.; writing---review and editing, M.H.E., N.O.P.V. and H.G.S.; supervision, B.D.N. and H.G.S. All authors have read and agreed to the published version of the manuscript.}

\funding{This research received no external funding.}

\institutionalreview{Not applicable.}

\informedconsent{Not applicable.}

\dataavailability{The synthetic datasets generated in this study are available from the corresponding author on reasonable request.}

\acknowledgments{We thank Francesca Azzolini for useful input on the statistical analysis.}

\conflictsofinterest{The authors declare no conflicts of interest.}

%=================================================================
\clearpage
\appendixstart
\appendix

\section{Scenario 1 result plots}
\label{app:scenario1_plots}
\appfigsection

Figure~\ref{fig:app_scenario1_grid} shows the full scatter plots underlying the $R^2$-values reported in Table~\ref{tab:scenario1_results} (Scenario 1, single SIE lens). Each row corresponds to one test (amplitudes truncated at 3rd, 4th, or 5th order); each column corresponds to one recovered lens parameter. In each panel, the found value is plotted against the ground truth, with the red line marking perfect recovery. For position, the blue (orange) points reference the position in the $\theta_1$ ($\theta_2$) direction.

\begin{figure}[p]
    \seedcontinuedfloat
    \centering
    \begin{center}
        \textbf{Test 1 ($\leq$3rd order)}
    \end{center}
    \begin{subfigure}{0.22\textwidth}
        \centering
        \includegraphics[width=\textwidth]{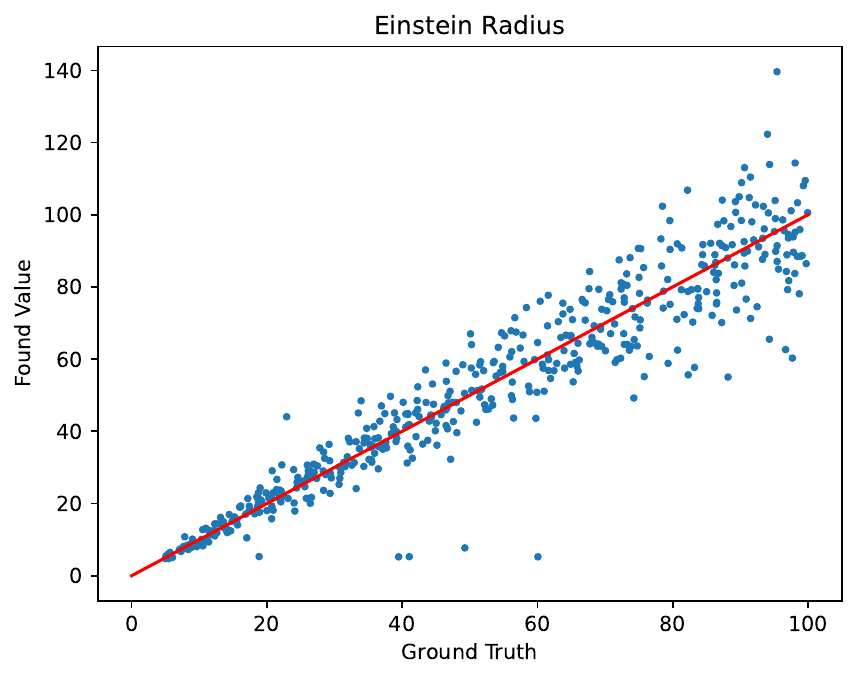}
    \end{subfigure}
    \begin{subfigure}{0.23\textwidth}
        \centering
        \includegraphics[width=\textwidth]{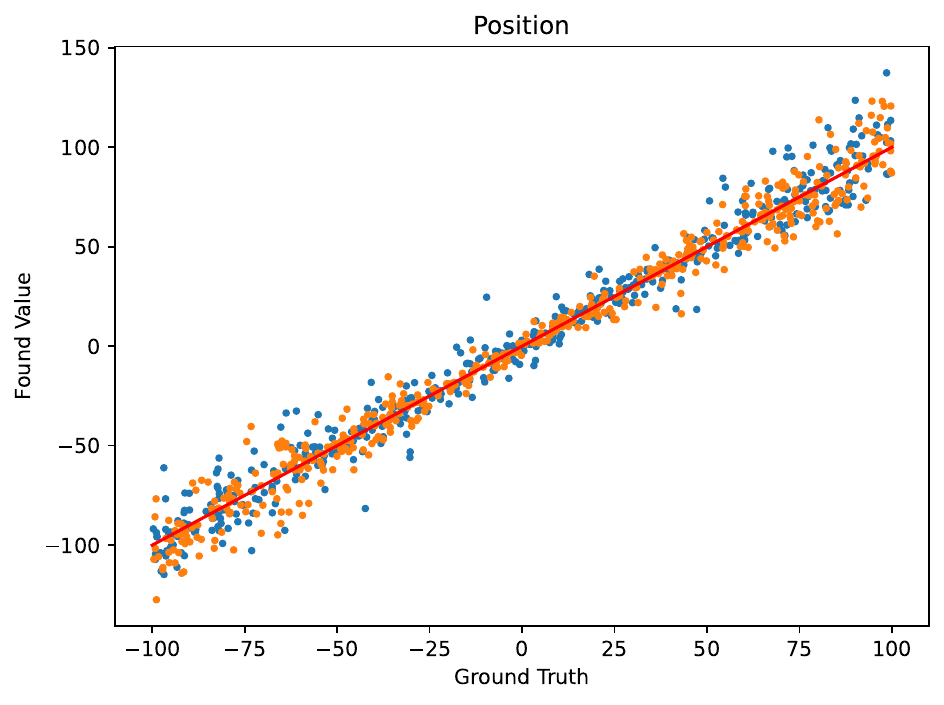}
    \end{subfigure}
    \begin{subfigure}{0.22\textwidth}
        \centering
        \includegraphics[width=\textwidth]{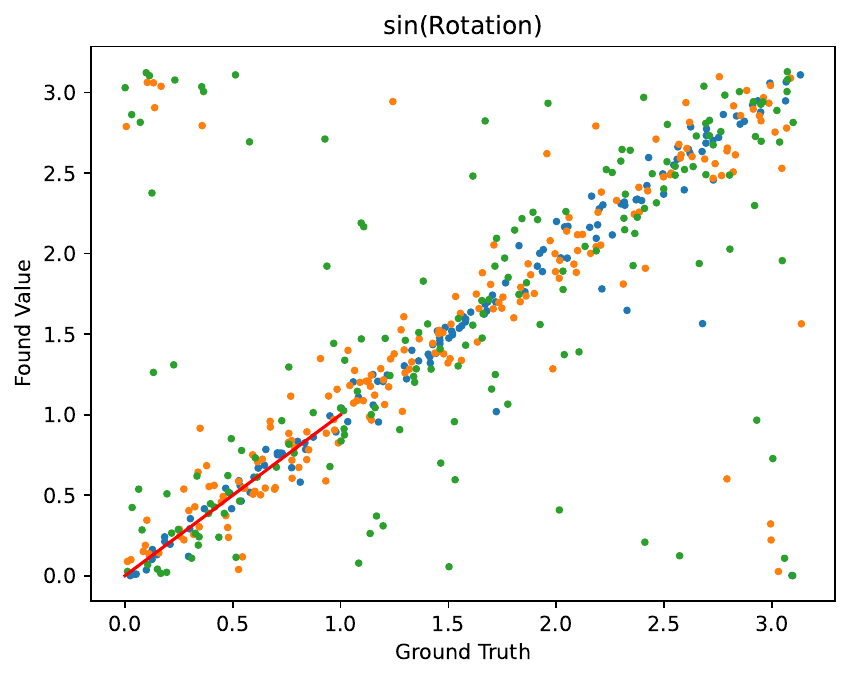}
    \end{subfigure}
    \begin{subfigure}{0.23\textwidth}
        \centering
        \includegraphics[width=\textwidth]{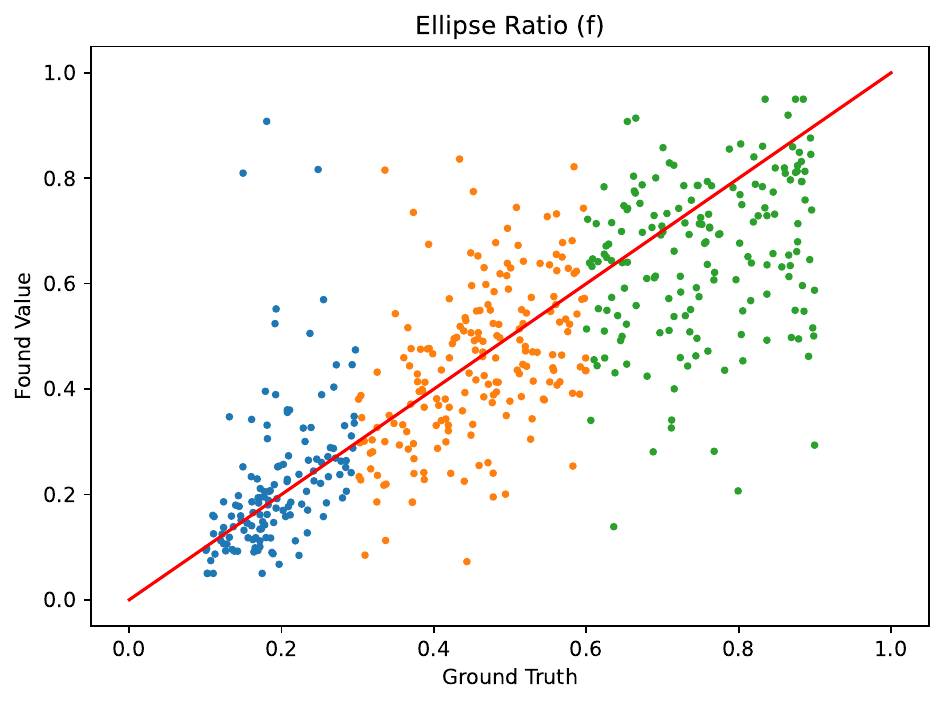}
    \end{subfigure}
\end{figure}

\begin{figure}[p]
    \ContinuedFloat
    \centering
    \begin{center}
        \textbf{Test 2 ($\leq$4th order)}
    \end{center}
    \begin{subfigure}{0.22\textwidth}
        \centering
        \includegraphics[width=\textwidth]{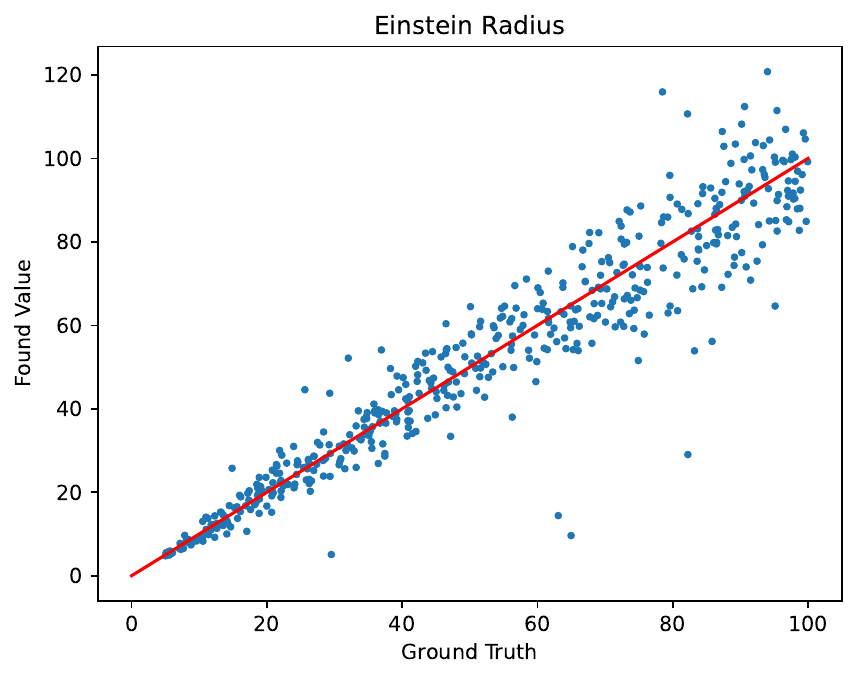}
    \end{subfigure}
    \begin{subfigure}{0.23\textwidth}
        \centering
        \includegraphics[width=\textwidth]{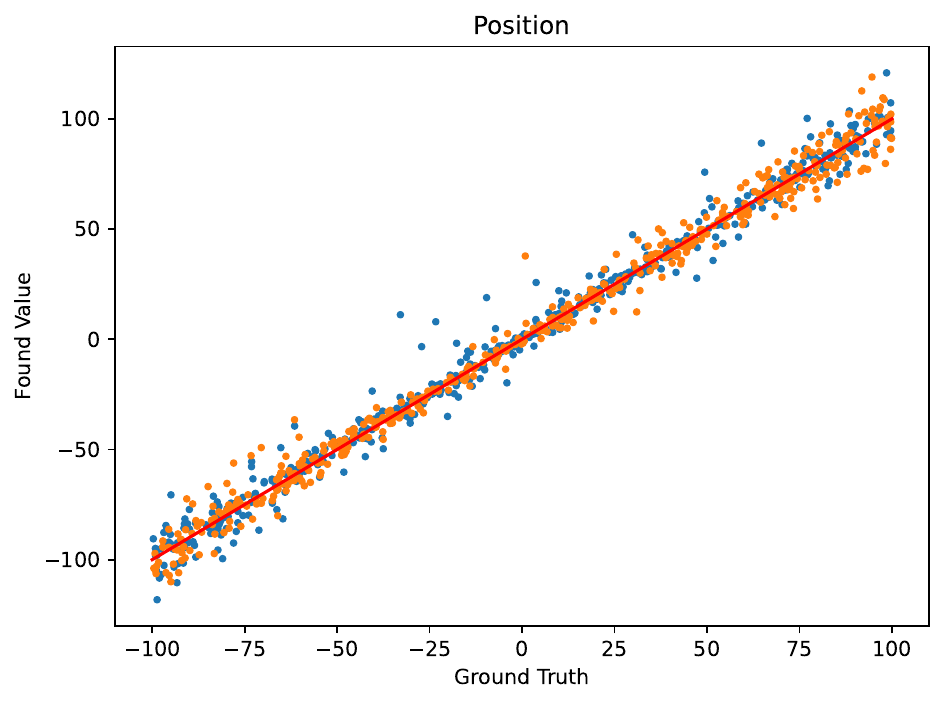}
    \end{subfigure}
    \begin{subfigure}{0.22\textwidth}
        \centering
        \includegraphics[width=\textwidth]{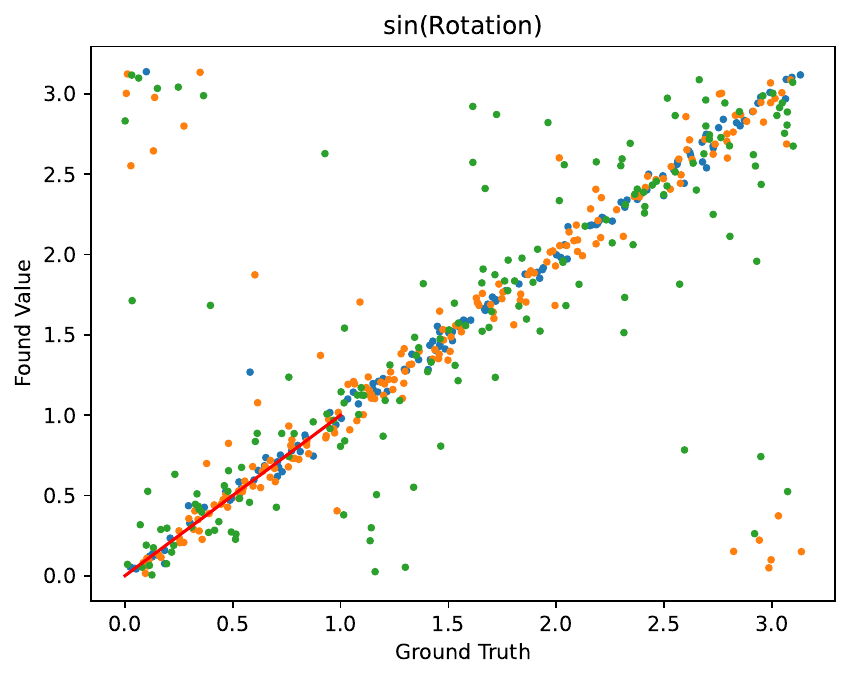}
    \end{subfigure}
    \begin{subfigure}{0.23\textwidth}
        \centering
        \includegraphics[width=\textwidth]{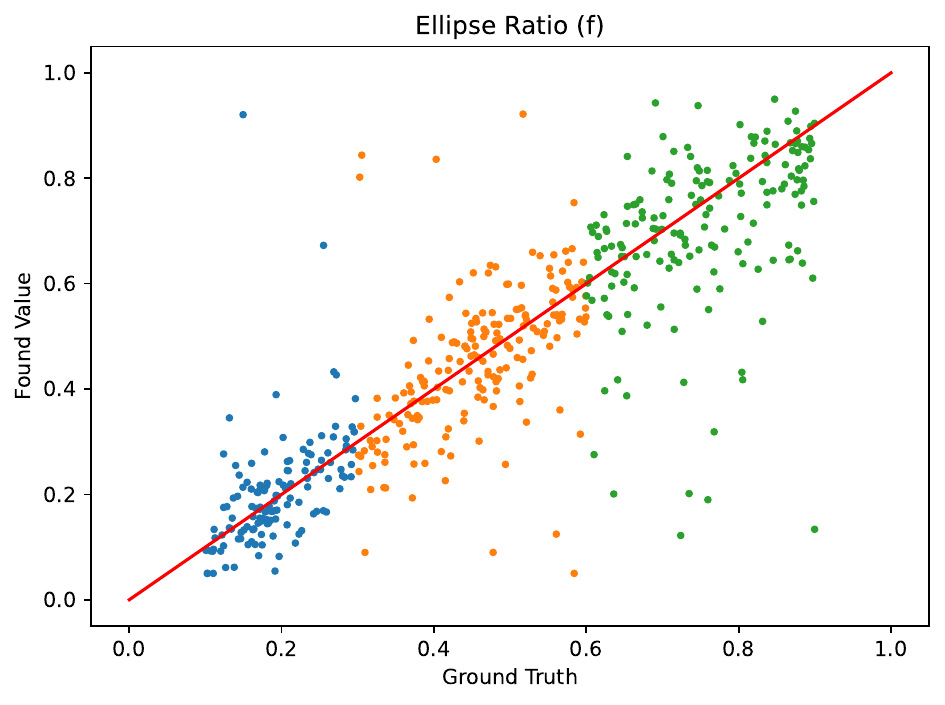}
    \end{subfigure}
\end{figure}

\begin{figure}[p]
    \ContinuedFloat
    \centering
    \begin{center}
        \textbf{Test 3 ($\leq$5th order)}
    \end{center}
    \begin{subfigure}{0.22\textwidth}
        \centering
        \includegraphics[width=\textwidth]{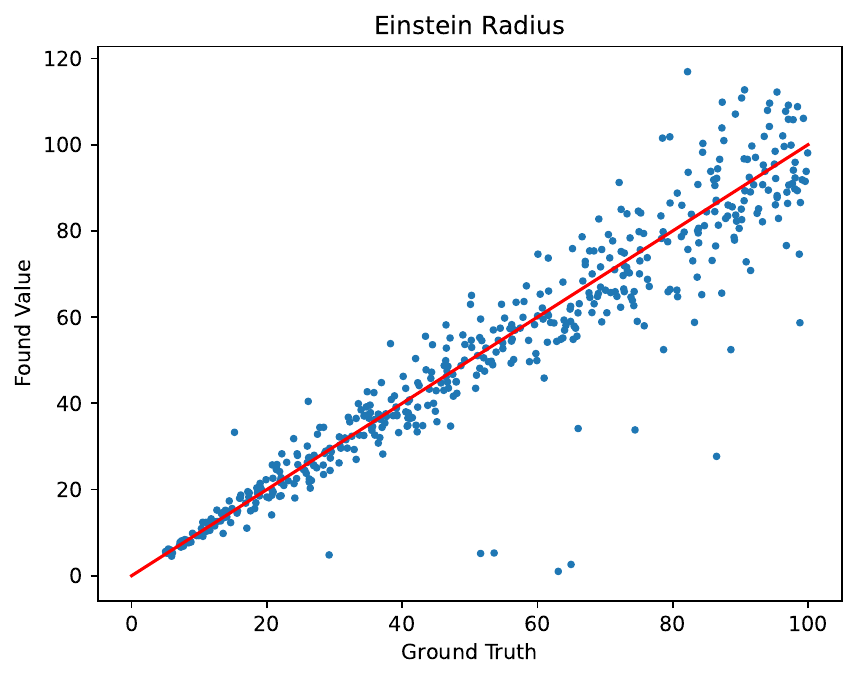}
    \end{subfigure}
    \begin{subfigure}{0.23\textwidth}
        \centering
        \includegraphics[width=\textwidth]{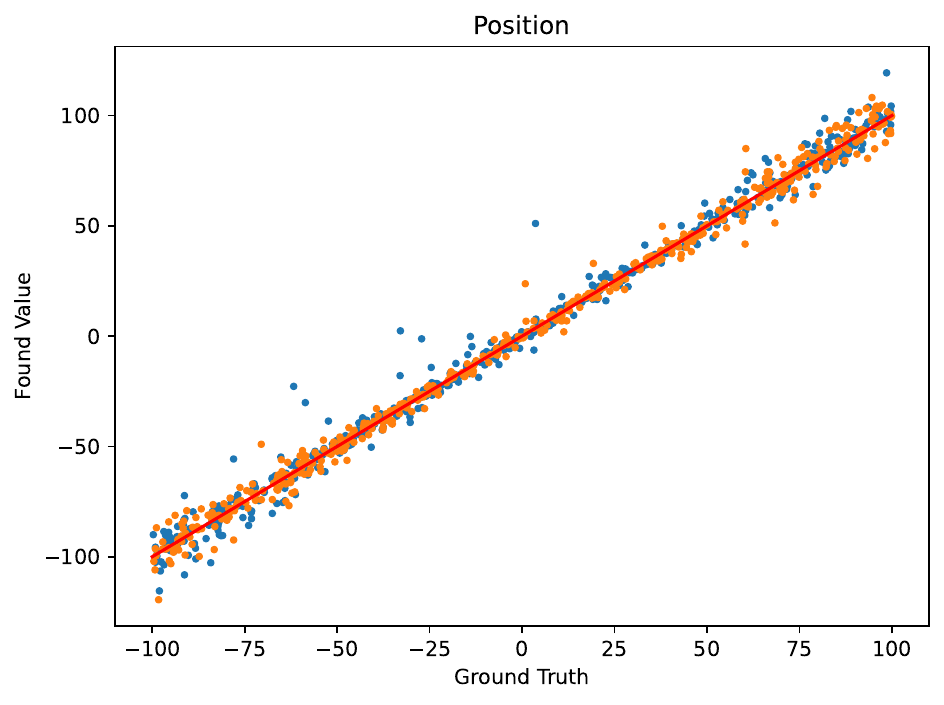}
    \end{subfigure}
    \begin{subfigure}{0.22\textwidth}
        \centering
        \includegraphics[width=\textwidth]{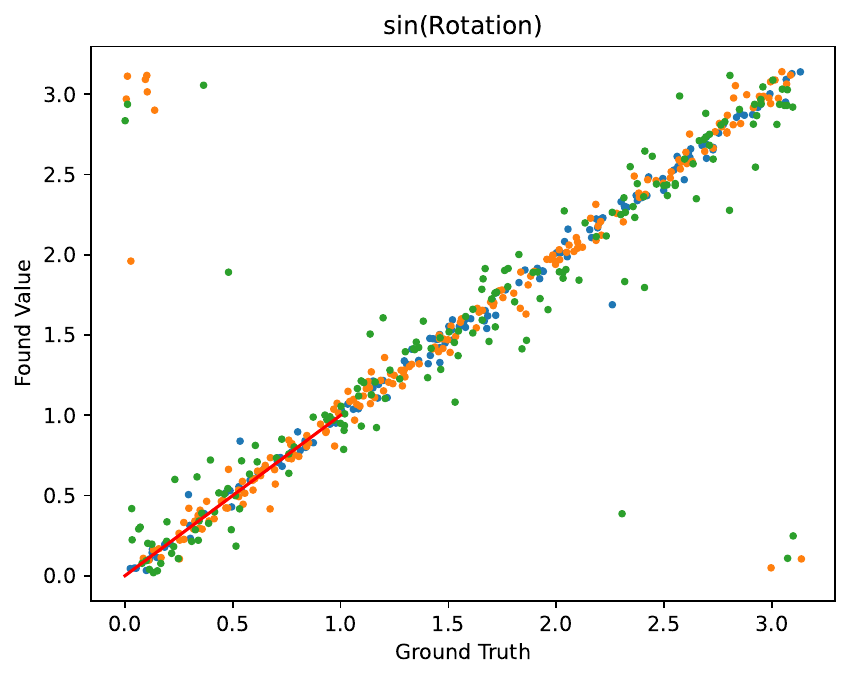}
    \end{subfigure}
    \begin{subfigure}{0.23\textwidth}
        \centering
        \includegraphics[width=\textwidth]{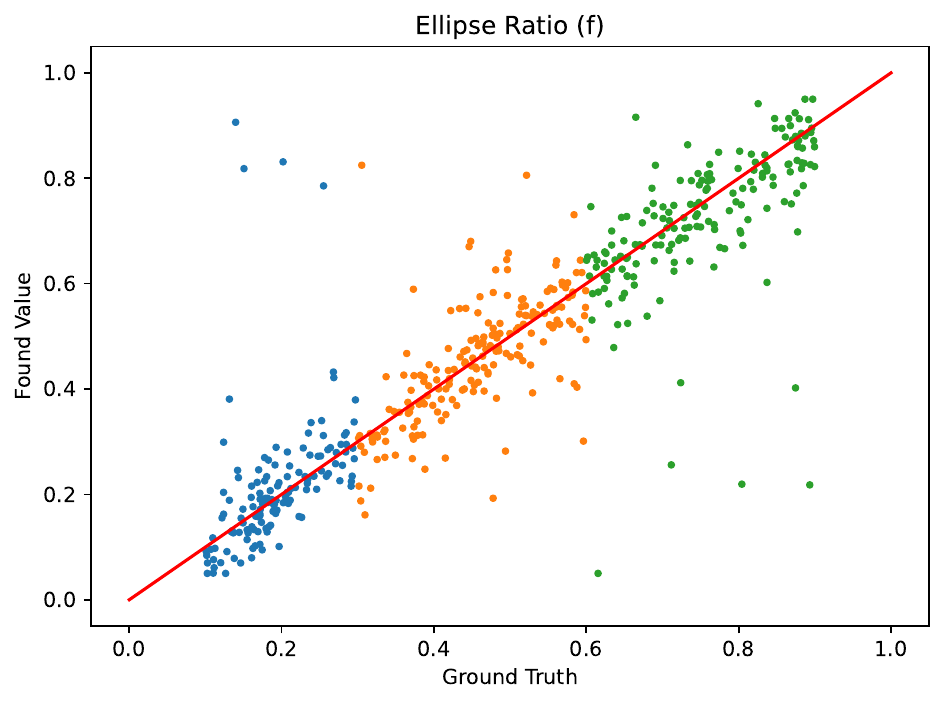}
    \end{subfigure}
    \appfigreset
    \caption{Scenario 1 (single SIE lens): scatter plots of found value versus ground truth for all three tests (rows) and all four lens parameters (columns: Einstein radius, position, sin(rotation), the ellipse ratio $f$). The red line marks the case where the found value equals the ground truth. For position, the blue (orange) points reference the position in the $\theta_1$ ($\theta_2$) direction. For the rotation and ellipse ratio, the lenses are color-coded depending on the value of the ellipse ratio for that lens. $f<0.3$ means it is blue, $0.3\leq f < 0.6$ means it is orange, otherwise it is green. The corresponding $R^2$-values are given in Table~\ref{tab:scenario1_results}.}
    \label{fig:app_scenario1_grid}
\end{figure}

\section{Scenario 2 result plots}
\label{app:scenario2_plots}
\appfigsection

Figures~\ref{fig:app_scenario2a_grid} and \ref{fig:app_scenario2b_grid} show the full scatter plots underlying the $R^2$-values reported in Table~\ref{tab:scenario2a_results} (Scenario 2a, SIE cluster, no noise) and Table~\ref{tab:scenario2b_results} (Scenario 2b, SIE cluster, with noise), respectively. Each row corresponds to one test (amplitudes truncated at 3rd, 4th, or 5th order); each column corresponds to one recovered lens parameter. In each panel, the found value is plotted against the ground truth, with the red line marking perfect recovery. For position, the blue (orange) points reference the position in the $\theta_1$ ($\theta_2$) direction.

\begin{figure}[p]
    \seedcontinuedfloat
    \centering
    \begin{center}
        \textbf{Test 1 ($\leq$3rd order)}
    \end{center}
    \begin{subfigure}{0.22\textwidth}
        \centering
        \includegraphics[width=\textwidth]{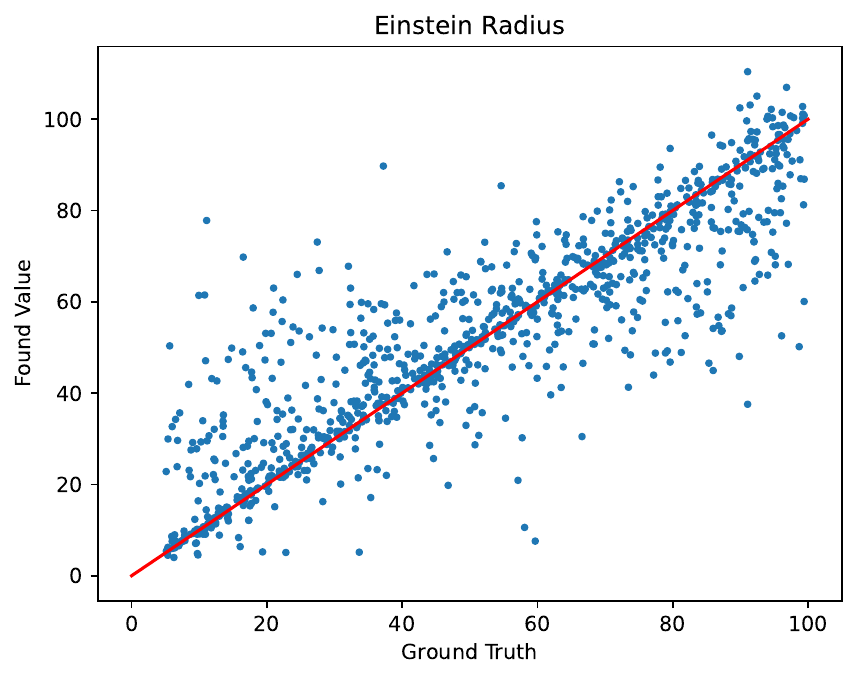}
    \end{subfigure}
    \begin{subfigure}{0.23\textwidth}
        \centering
        \includegraphics[width=\textwidth]{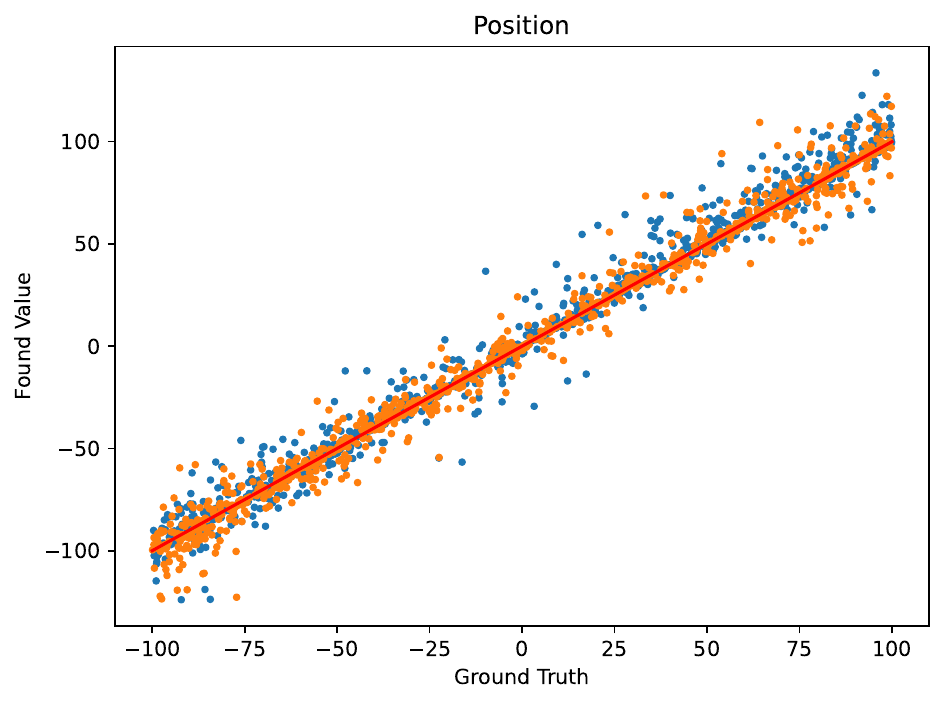}
    \end{subfigure}
    \begin{subfigure}{0.22\textwidth}
        \centering
        \includegraphics[width=\textwidth]{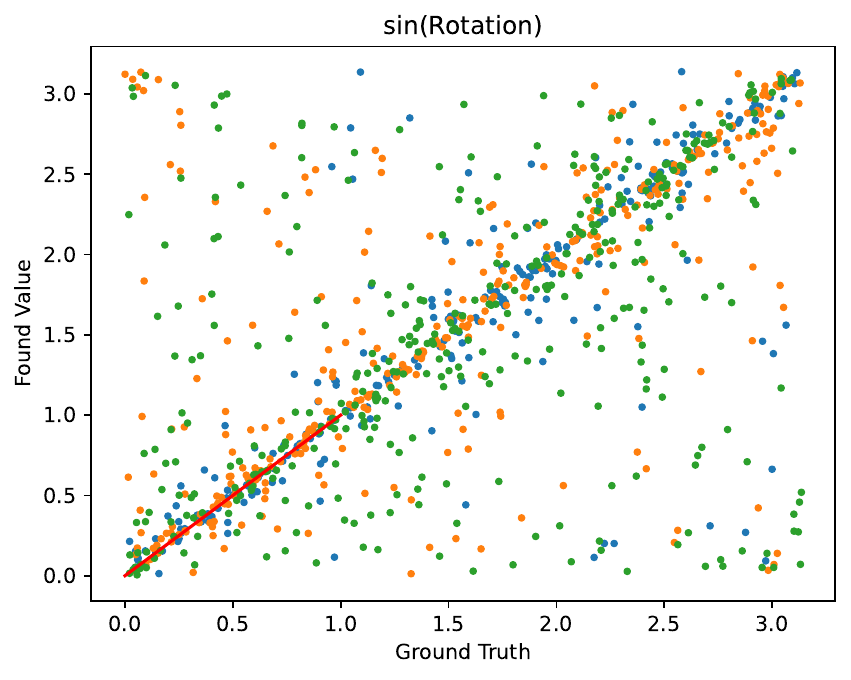}
    \end{subfigure}
    \begin{subfigure}{0.23\textwidth}
        \centering
        \includegraphics[width=\textwidth]{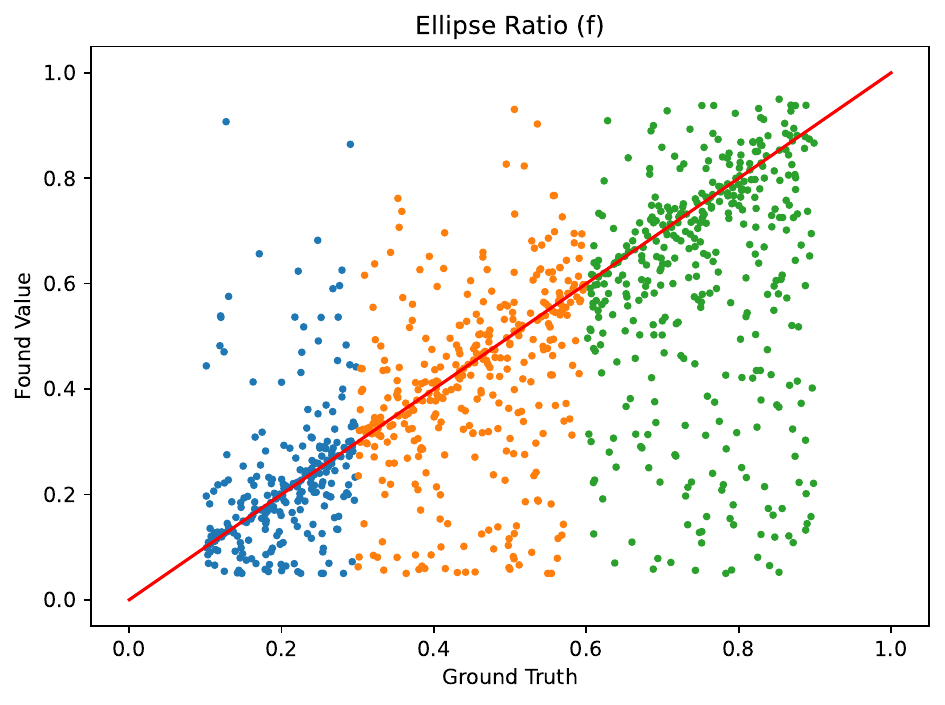}
    \end{subfigure}
\end{figure}

\begin{figure}[p]
    \ContinuedFloat
    \centering
    \begin{center}
        \textbf{Test 2 ($\leq$4th order)}
    \end{center}
    \begin{subfigure}{0.22\textwidth}
        \centering
        \includegraphics[width=\textwidth]{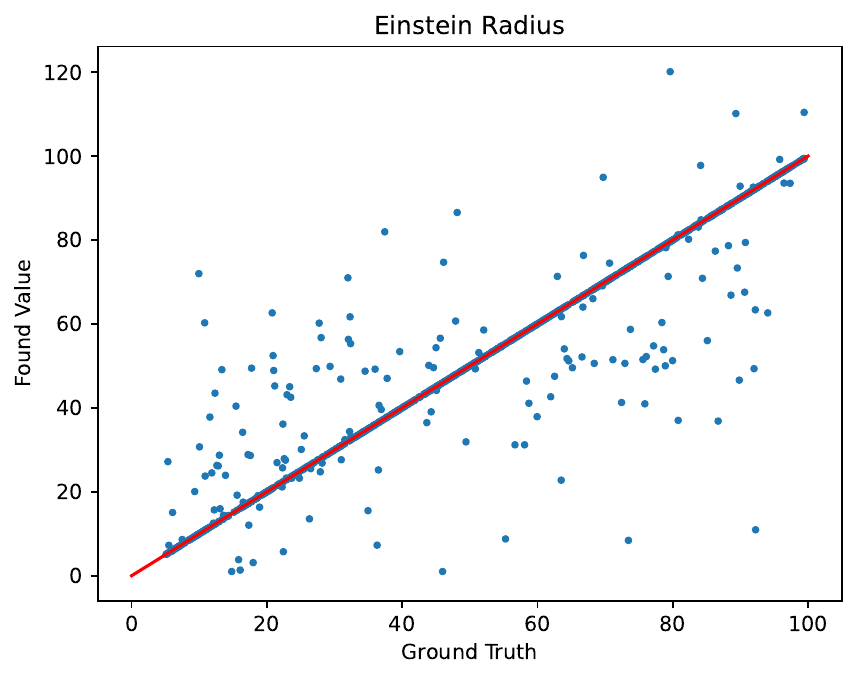}
    \end{subfigure}
    \begin{subfigure}{0.23\textwidth}
        \centering
        \includegraphics[width=\textwidth]{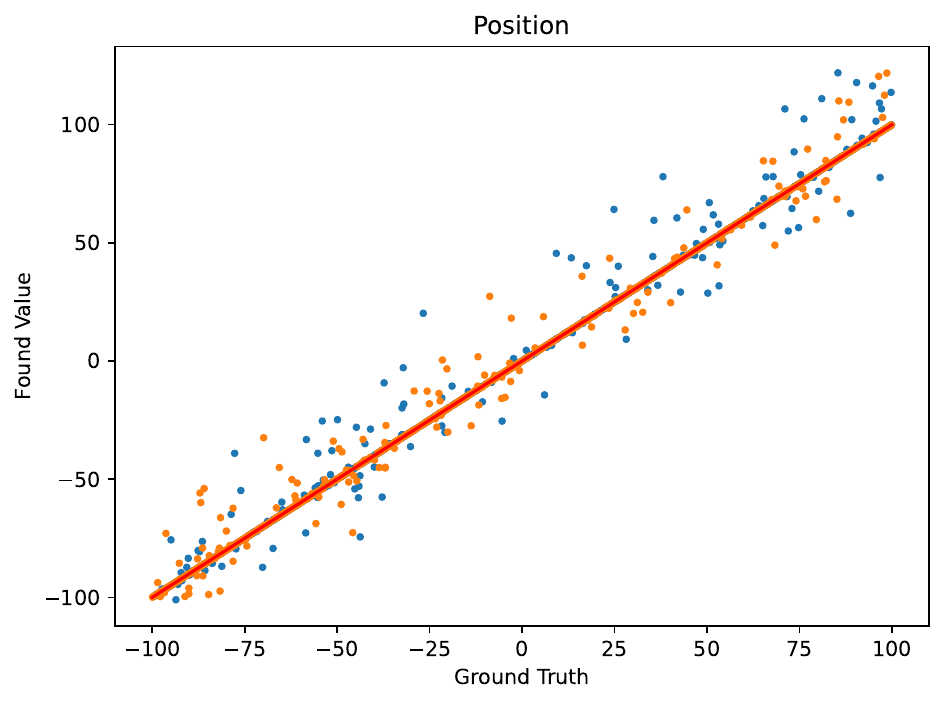}
    \end{subfigure}
    \begin{subfigure}{0.22\textwidth}
        \centering
        \includegraphics[width=\textwidth]{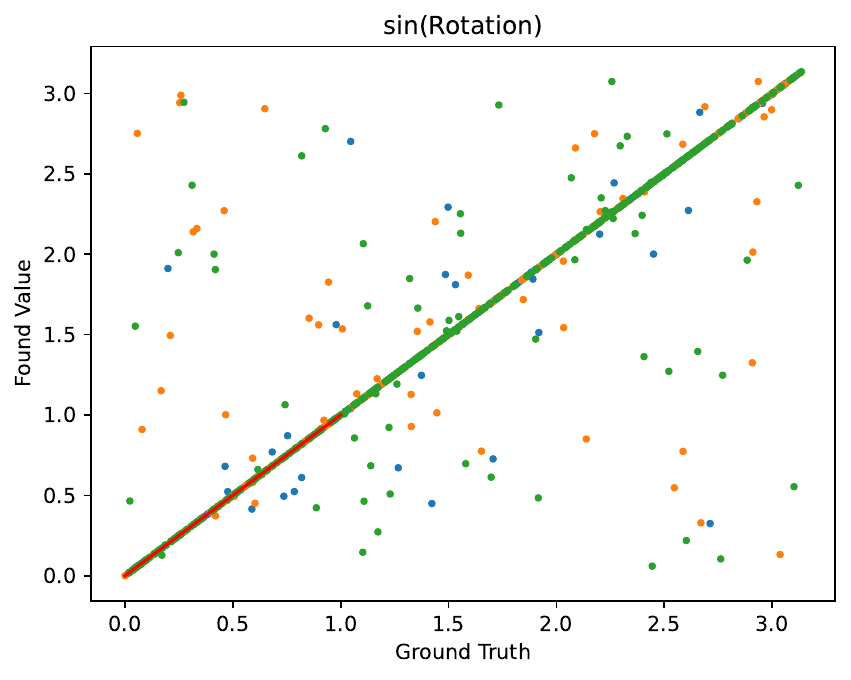}
    \end{subfigure}
    \begin{subfigure}{0.23\textwidth}
        \centering
        \includegraphics[width=\textwidth]{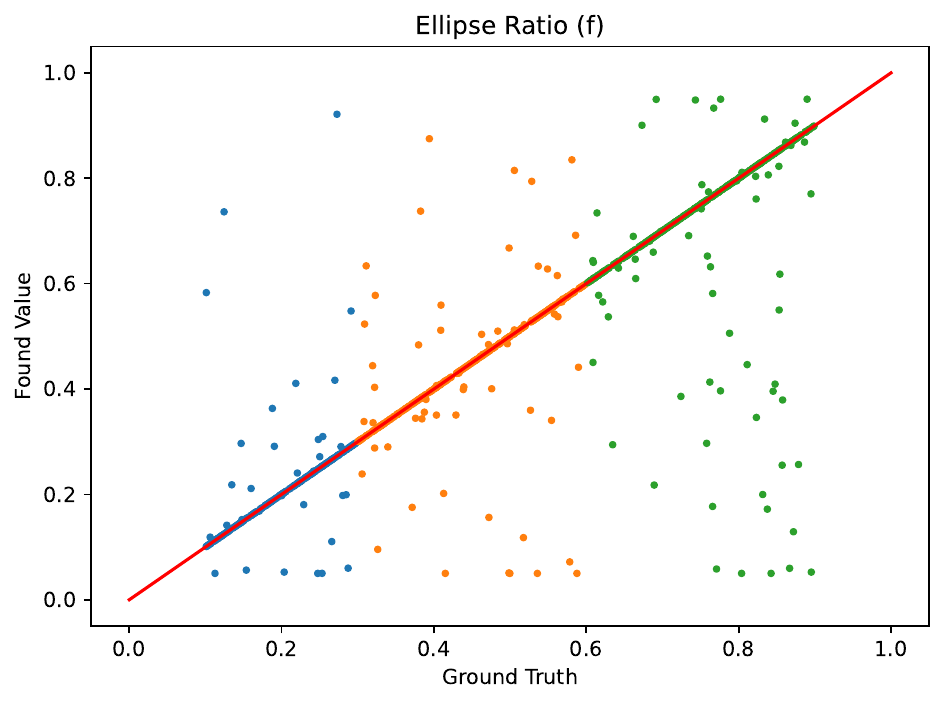}
    \end{subfigure}
\end{figure}

\begin{figure}[p]
    \ContinuedFloat
    \centering
    \begin{center}
        \textbf{Test 3 ($\leq$5th order)}
    \end{center}
    \begin{subfigure}{0.22\textwidth}
        \centering
        \includegraphics[width=\textwidth]{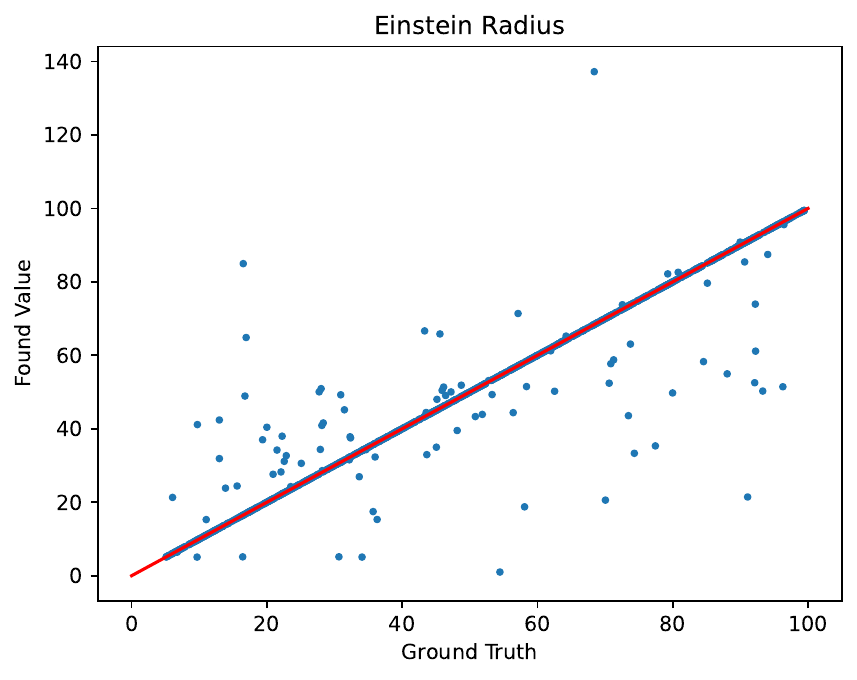}
    \end{subfigure}
    \begin{subfigure}{0.23\textwidth}
        \centering
        \includegraphics[width=\textwidth]{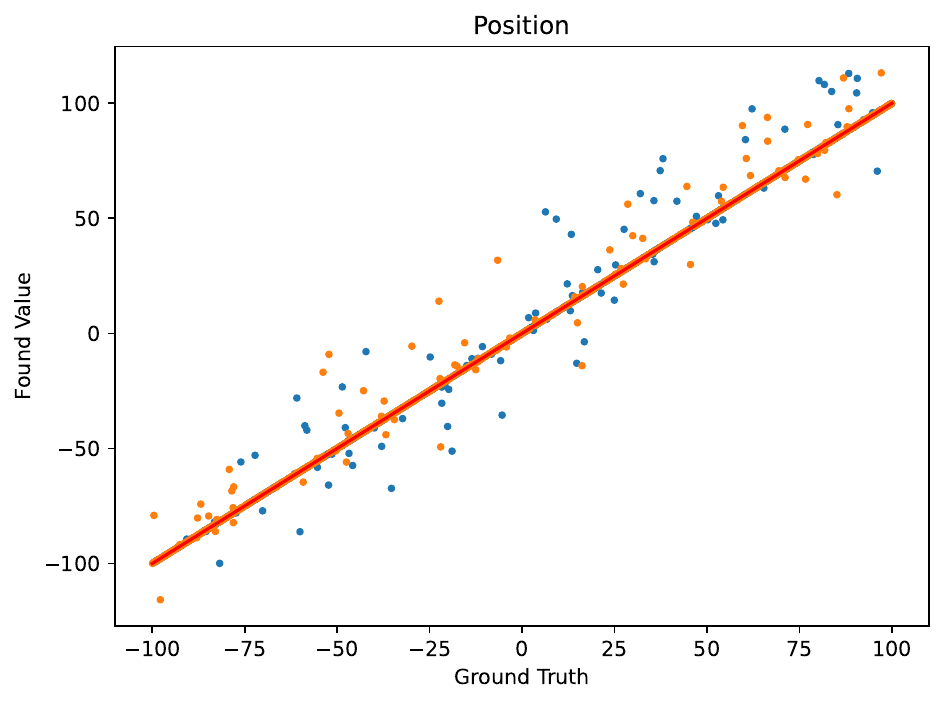}
    \end{subfigure}
    \begin{subfigure}{0.22\textwidth}
        \centering
        \includegraphics[width=\textwidth]{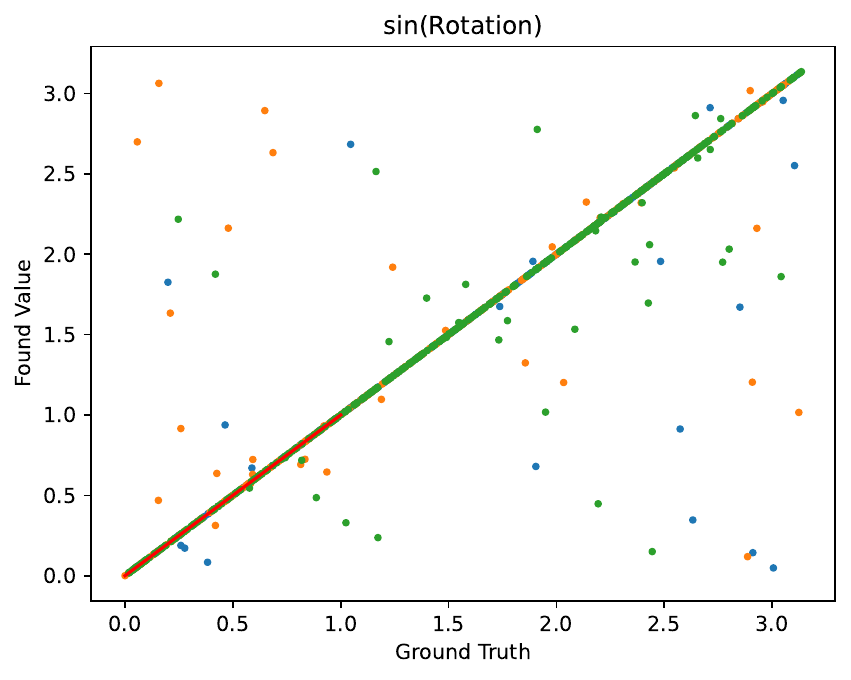}
    \end{subfigure}
    \begin{subfigure}{0.23\textwidth}
        \centering
        \includegraphics[width=\textwidth]{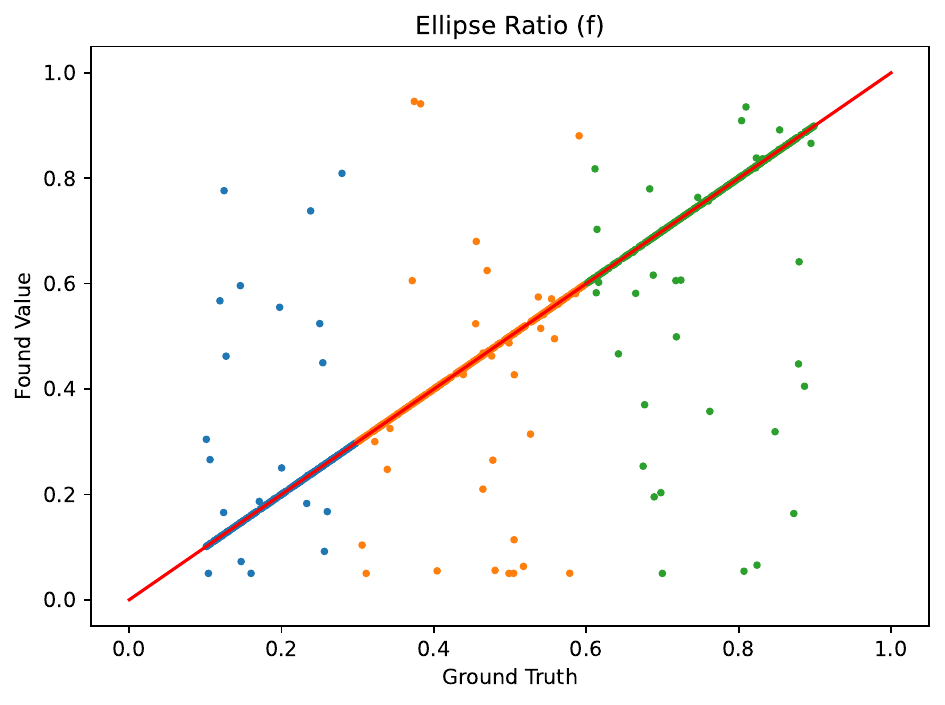}
    \end{subfigure}
    \appfigreset
    \caption{Scenario 2a (SIE cluster lens, no noise): scatter plots of found value versus ground truth for all three tests (rows) and all four lens parameters (columns: Einstein radius, position, sin(rotation), the ellipse ratio $f$). The red line marks the case where the found value equals the ground truth. For position, the blue (orange) points reference the position in the $\theta_1$ ($\theta_2$) direction. For the rotation and ellipse ratio, the lenses are color-coded depending on the value of the ellipse ratio for that lens. $f<0.3$ means it is blue, $0.3\leq f < 0.6$ means it is orange, otherwise it is green. The corresponding $R^2$-values are given in Table~\ref{tab:scenario2a_results}.}
    \label{fig:app_scenario2a_grid}
\end{figure}

\clearpage
\begin{figure}[p]
    \seedcontinuedfloat
    \centering
    \begin{center}
        \textbf{Test 1 ($\leq$3rd order)}
    \end{center}
    \begin{subfigure}{0.22\textwidth}
        \centering
        \includegraphics[width=\textwidth]{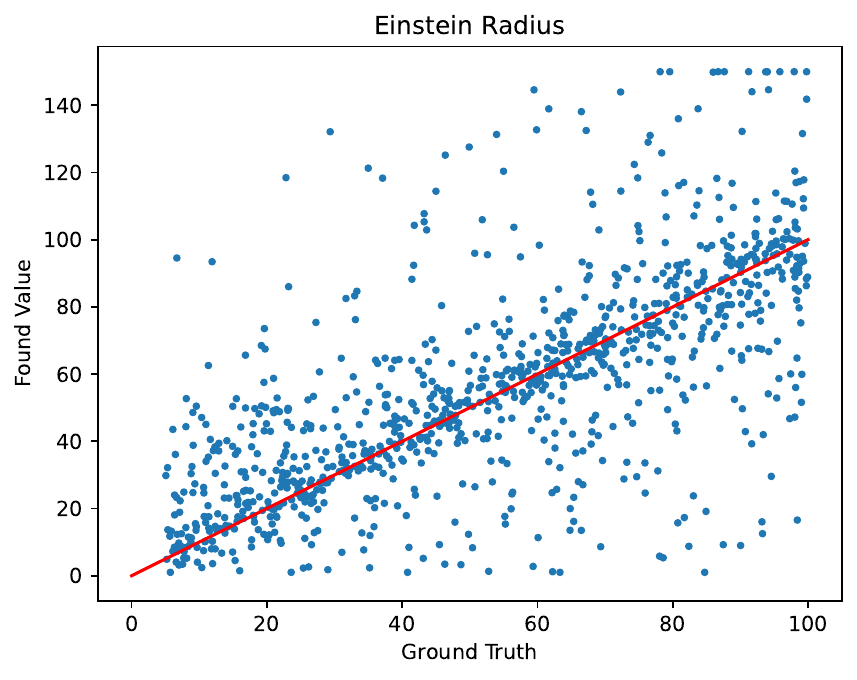}
    \end{subfigure}
    \begin{subfigure}{0.23\textwidth}
        \centering
        \includegraphics[width=\textwidth]{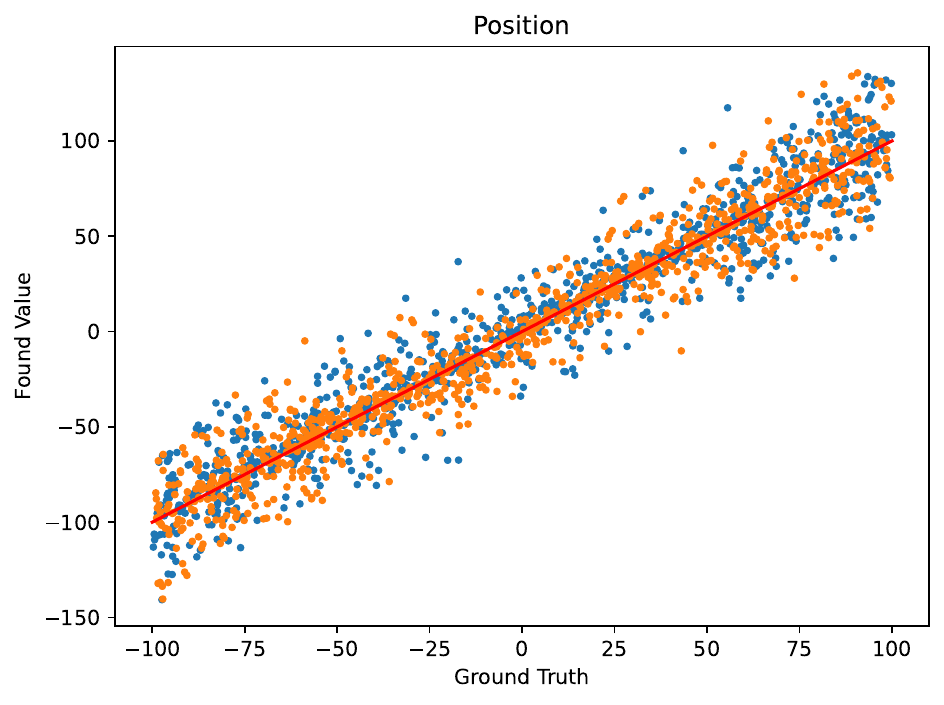}
    \end{subfigure}
    \begin{subfigure}{0.22\textwidth}
        \centering
        \includegraphics[width=\textwidth]{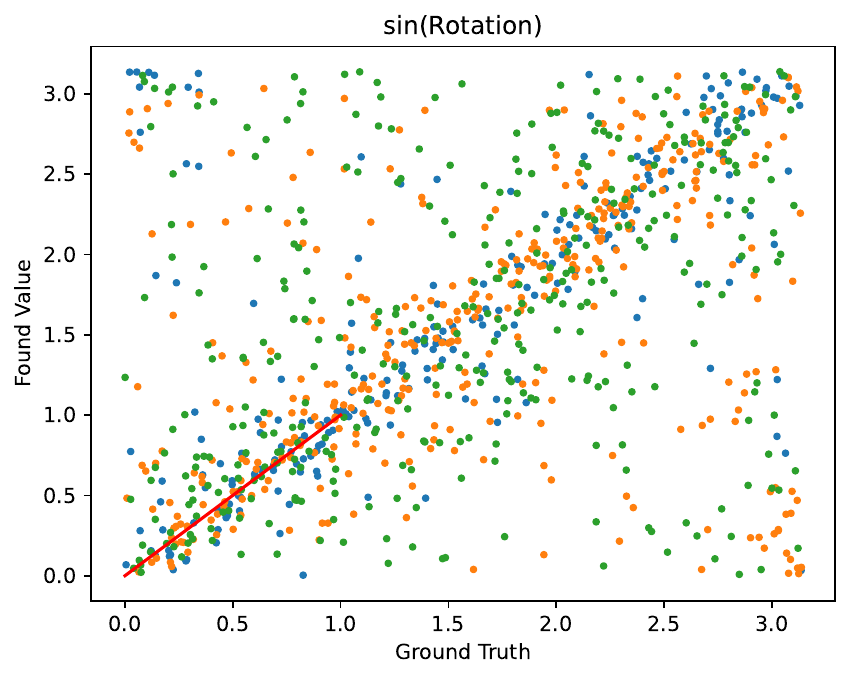}
    \end{subfigure}
    \begin{subfigure}{0.23\textwidth}
        \centering
        \includegraphics[width=\textwidth]{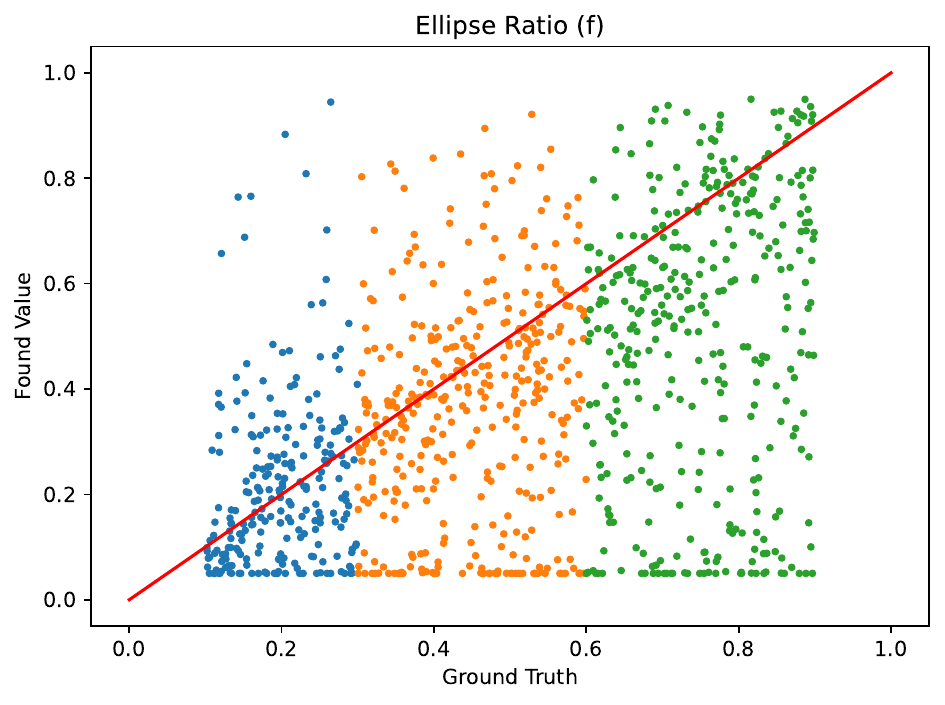}
    \end{subfigure}
\end{figure}

\begin{figure}[p]
    \ContinuedFloat
    \centering
    \begin{center}
        \textbf{Test 2 ($\leq$4th order)}
    \end{center}
    \begin{subfigure}{0.22\textwidth}
        \centering
        \includegraphics[width=\textwidth]{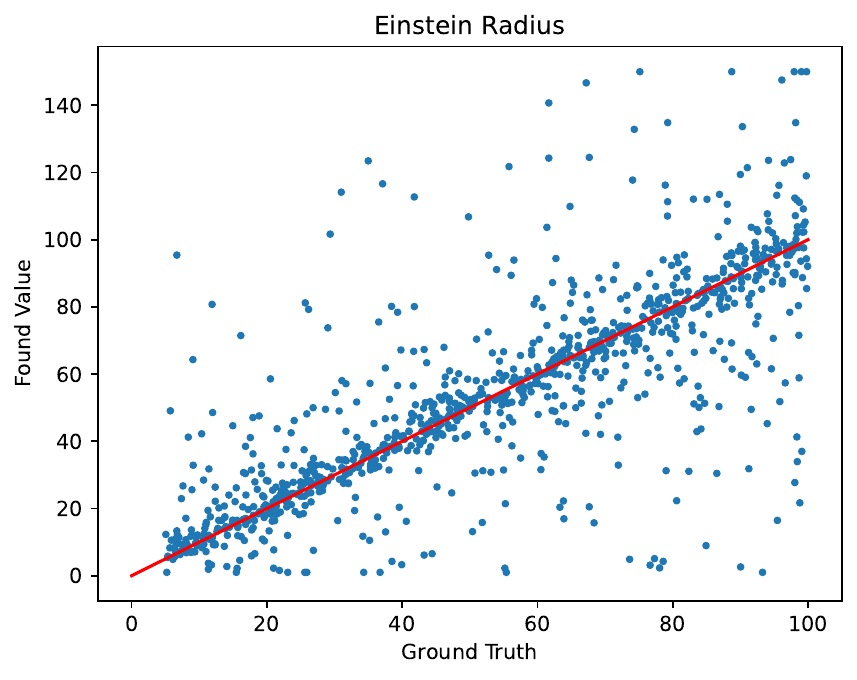}
    \end{subfigure}
    \begin{subfigure}{0.23\textwidth}
        \centering
        \includegraphics[width=\textwidth]{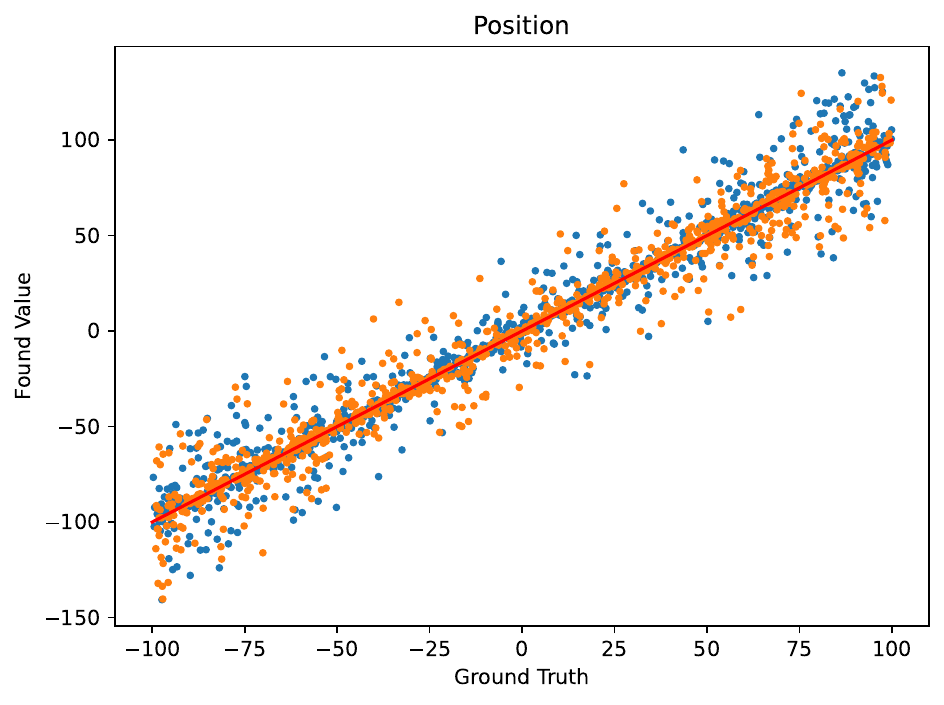}
    \end{subfigure}
    \begin{subfigure}{0.22\textwidth}
        \centering
        \includegraphics[width=\textwidth]{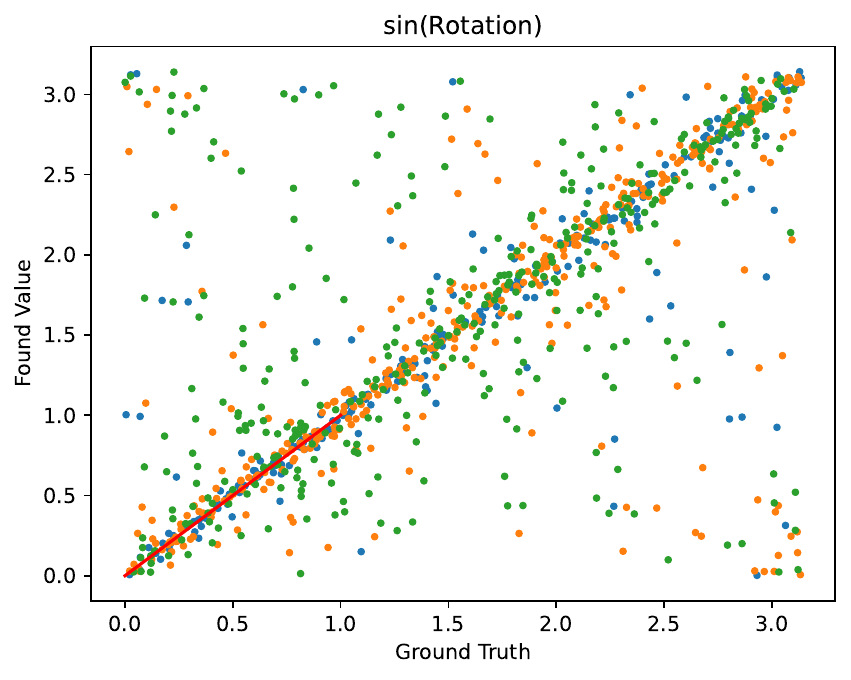}
    \end{subfigure}
    \begin{subfigure}{0.23\textwidth}
        \centering
        \includegraphics[width=\textwidth]{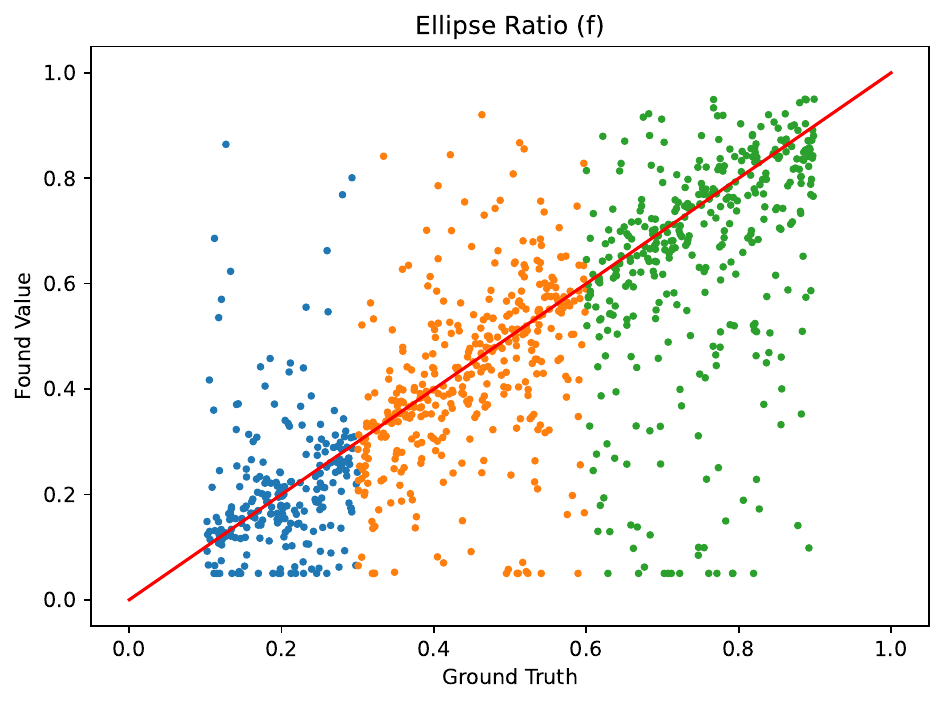}
    \end{subfigure}
\end{figure}

\begin{figure}[p]
    \ContinuedFloat
    \centering
    \begin{center}
        \textbf{Test 3 ($\leq$5th order)}
    \end{center}
    \begin{subfigure}{0.22\textwidth}
        \centering
        \includegraphics[width=\textwidth]{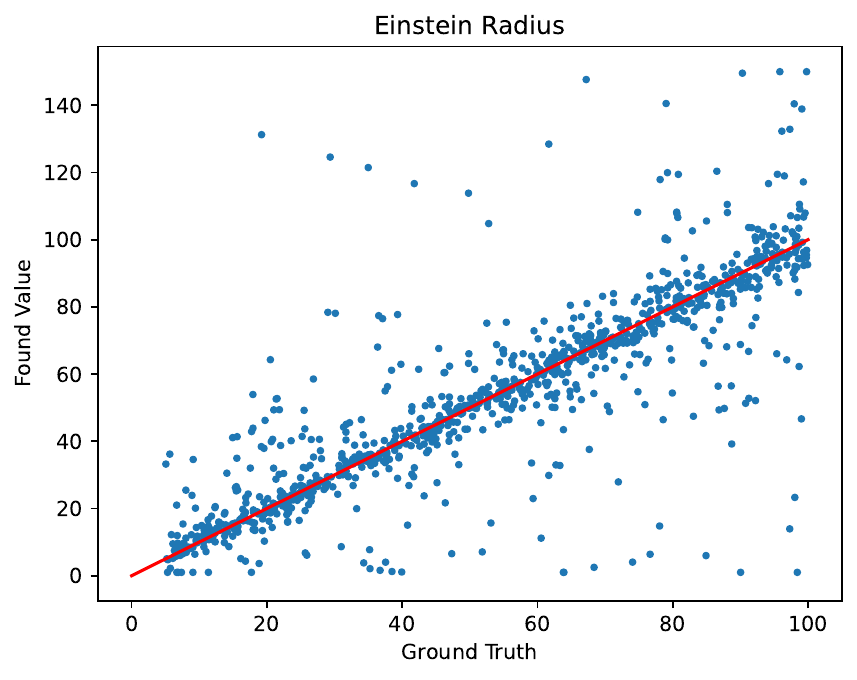}
    \end{subfigure}
    \begin{subfigure}{0.23\textwidth}
        \centering
        \includegraphics[width=\textwidth]{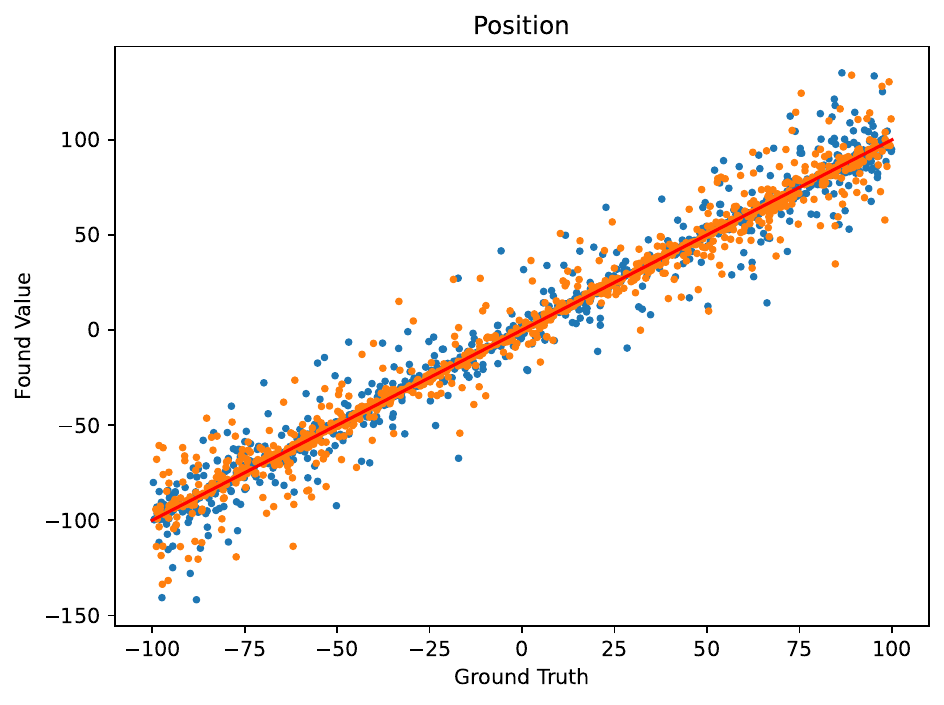}
    \end{subfigure}
    \begin{subfigure}{0.22\textwidth}
        \centering
        \includegraphics[width=\textwidth]{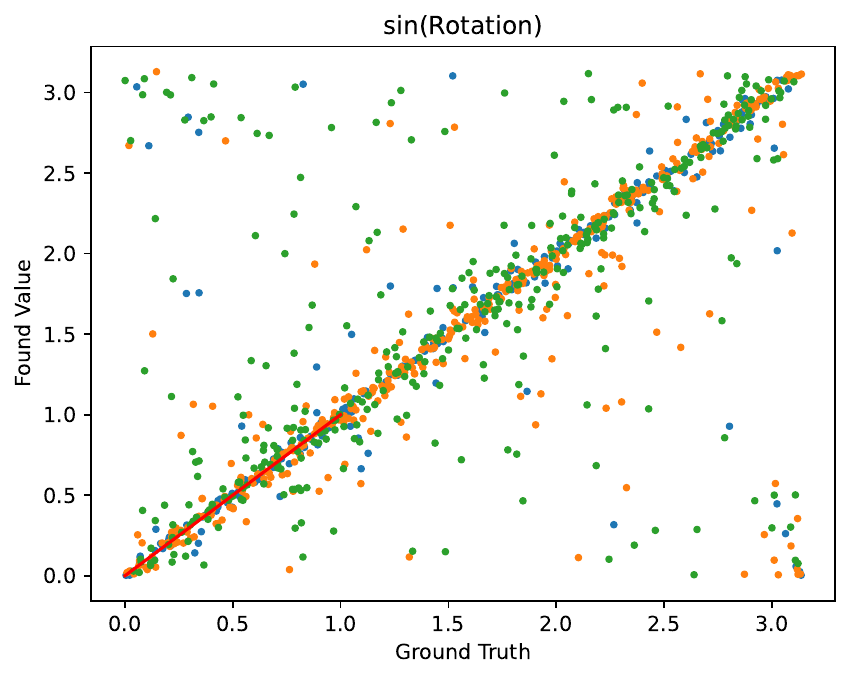}
    \end{subfigure}
    \begin{subfigure}{0.23\textwidth}
        \centering
        \includegraphics[width=\textwidth]{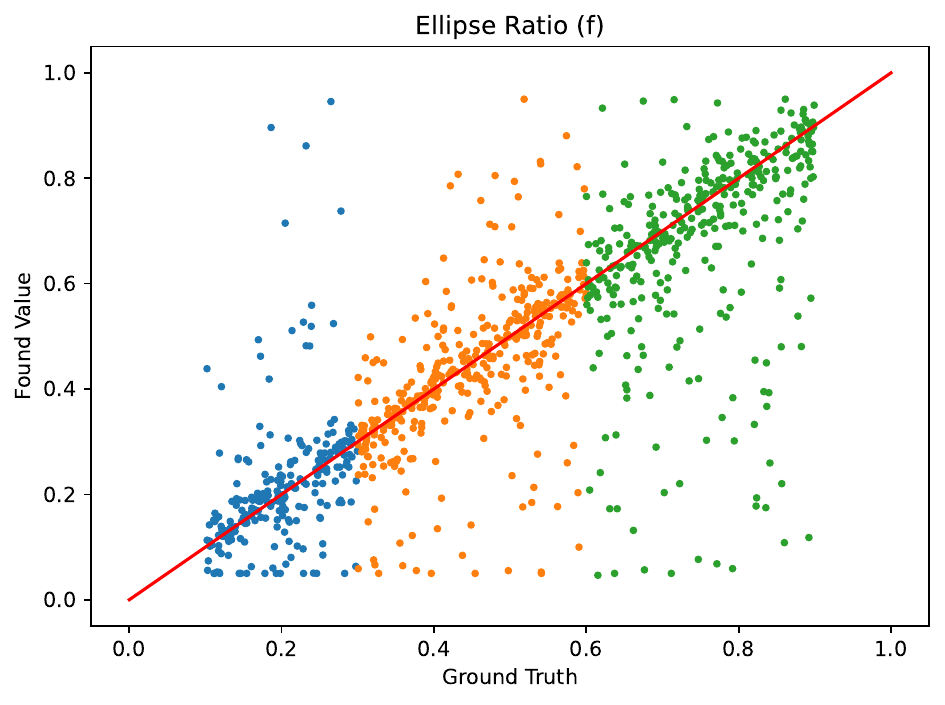}
    \end{subfigure}
    \appfigreset
    \caption{Scenario 2b (SIE cluster lens, with noise): scatter plots of found value versus ground truth for all three tests (rows) and all four lens parameters (columns: Einstein radius, position, sin(rotation), the ellipse ratio $f$). The red line marks the case where the found value equals the ground truth. For position, the blue (orange) points reference the position in the $\theta_1$ ($\theta_2$) direction. For the rotation and ellipse ratio, the lenses are color-coded depending on the value of the ellipse ratio for that lens. $f<0.3$ means it is blue, $0.3\leq f < 0.6$ means it is orange, otherwise it is green. The corresponding $R^2$-values are given in Table~\ref{tab:scenario2b_results}.}
    \label{fig:app_scenario2b_grid}
\end{figure}

\section{Scenario 3 result plots}
\label{app:scenario3_plots}
\appfigsection

Figures~\ref{fig:app_scenario3a_grid} and \ref{fig:app_scenario3b_grid} show the full scatter plots underlying the $R^2$-values reported in Table~\ref{tab:scenario3a_results} (Scenario 3a, SIS cluster, no noise) and Table~\ref{tab:scenario3b_results} (Scenario 3b, SIS cluster, with noise), respectively. Each row corresponds to one test (amplitudes truncated at 3rd, 4th, or 5th order); each column corresponds to one recovered lens parameter. In each panel, the found value is plotted against the ground truth, with the red line marking perfect recovery. For position, the blue (orange) points reference the position in the $\theta_1$ ($\theta_2$) direction.

\begin{figure}[p]
    \seedcontinuedfloat
    \centering
    \begin{center}
        \textbf{Test 1 ($\leq$3rd order)}
    \end{center}
    \begin{subfigure}{0.4\textwidth}
        \centering
        \includegraphics[width=\textwidth]{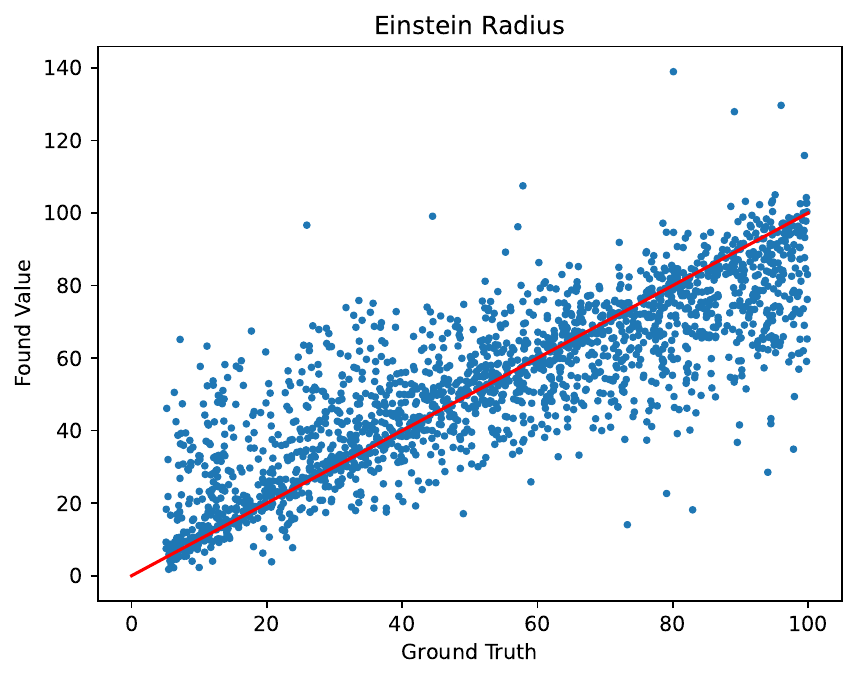}
    \end{subfigure}
    \begin{subfigure}{0.415\textwidth}
        \centering
        \includegraphics[width=\textwidth]{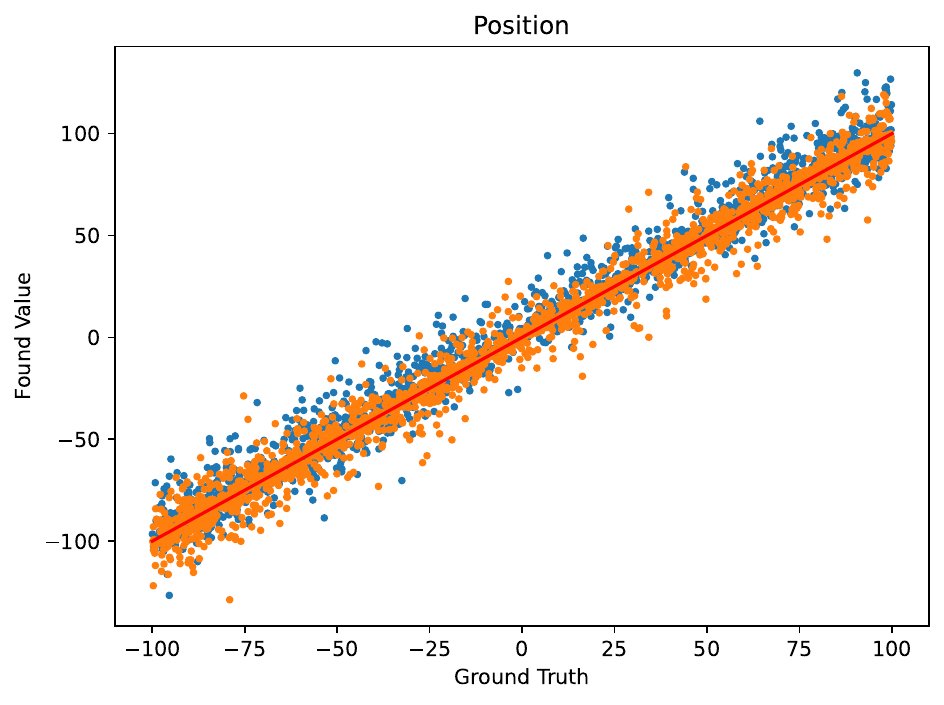}
    \end{subfigure}
\end{figure}

\begin{figure}[p]
    \ContinuedFloat
    \centering
    \begin{center}
        \textbf{Test 2 ($\leq$4th order)}
    \end{center}
    \begin{subfigure}{0.4\textwidth}
        \centering
        \includegraphics[width=\textwidth]{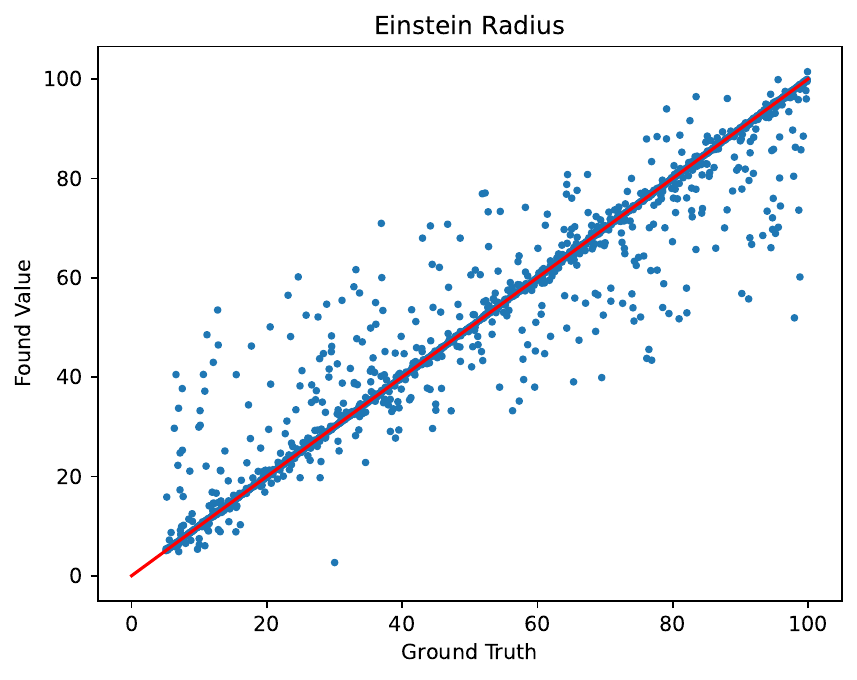}
    \end{subfigure}
    \begin{subfigure}{0.415\textwidth}
        \centering
        \includegraphics[width=\textwidth]{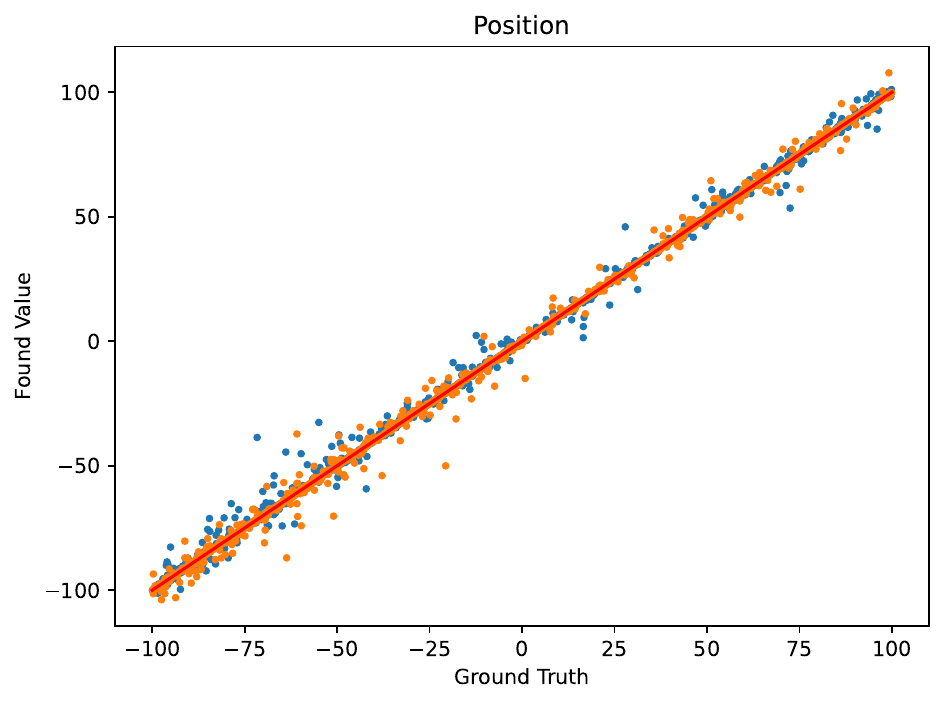}
    \end{subfigure}
\end{figure}

\begin{figure}[p]
    \ContinuedFloat
    \centering
    \begin{center}
        \textbf{Test 3 ($\leq$5th order)}
    \end{center}
    \begin{subfigure}{0.4\textwidth}
        \centering
        \includegraphics[width=\textwidth]{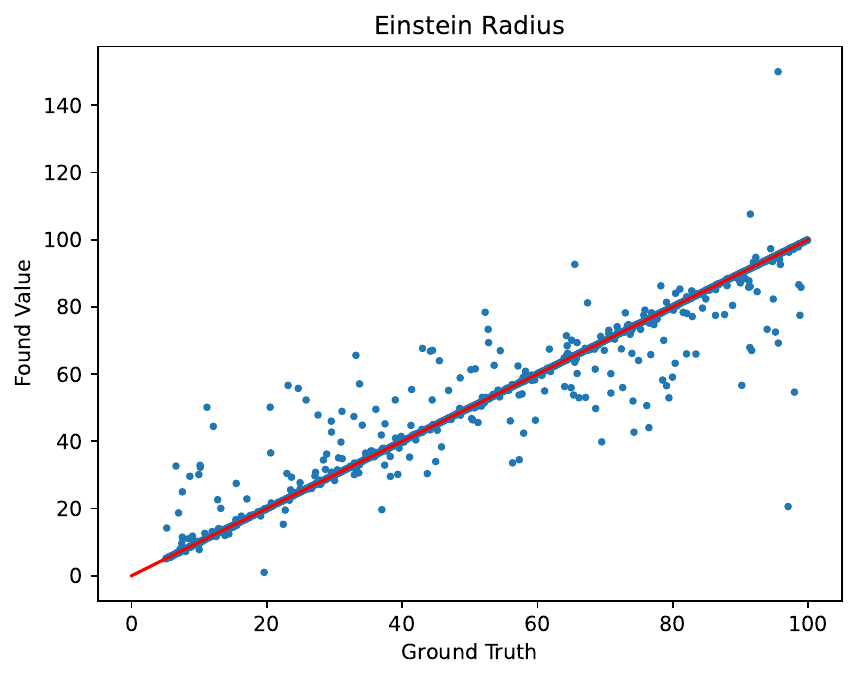}
    \end{subfigure}
    \begin{subfigure}{0.415\textwidth}
        \centering
        \includegraphics[width=\textwidth]{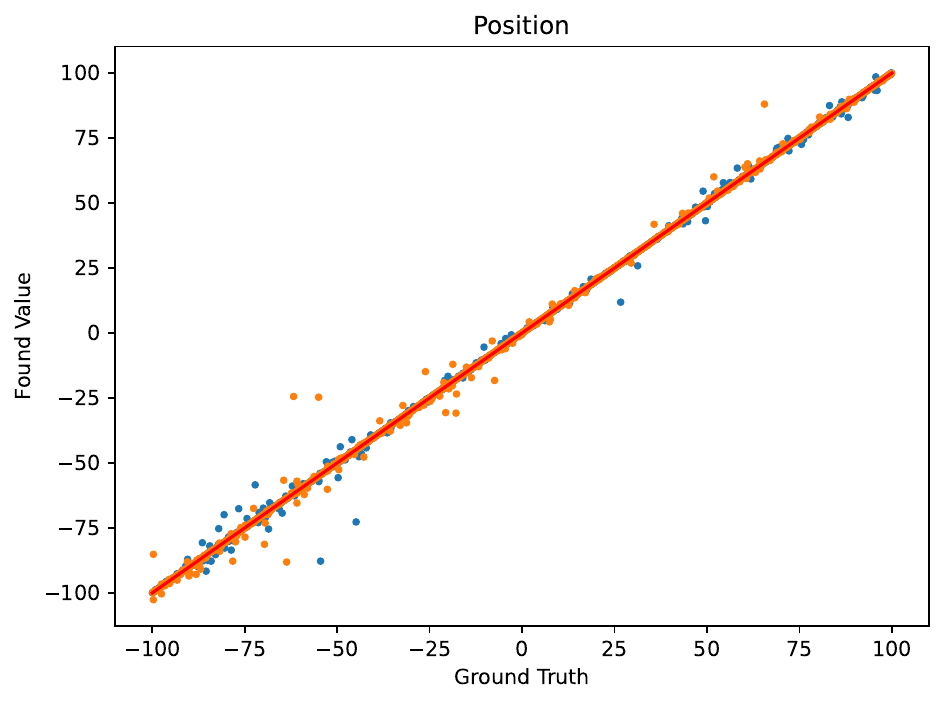}
    \end{subfigure}
    \appfigreset
    \caption{Scenario 3a (SIS cluster lens, no noise): scatter plots of found value versus ground truth for all three tests (rows) and both recovered lens parameters (columns: Einstein radius, position). The red line marks the case where the found value equals the ground truth. For position, the blue (orange) points reference the position in the $\theta_1$ ($\theta_2$) direction. The corresponding $R^2$-values are given in Table~\ref{tab:scenario3a_results}.}
    \label{fig:app_scenario3a_grid}
\end{figure}

\clearpage
\begin{figure}[p]
    \seedcontinuedfloat
    \centering
    \begin{center}
        \textbf{Test 1 ($\leq$3rd order)}
    \end{center}
    \begin{subfigure}{0.4\textwidth}
        \centering
        \includegraphics[width=\textwidth]{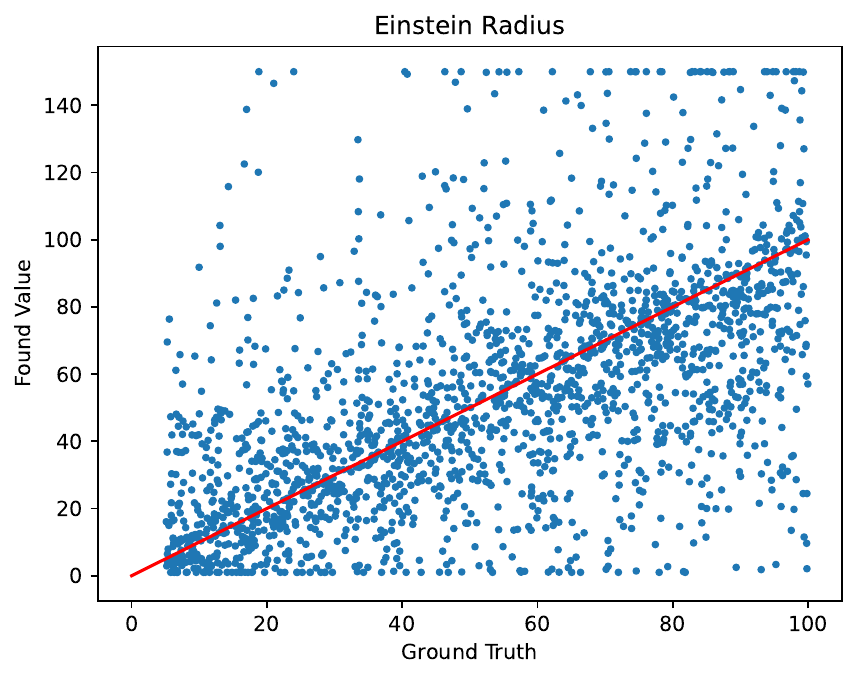}
    \end{subfigure}
    \begin{subfigure}{0.415\textwidth}
        \centering
        \includegraphics[width=\textwidth]{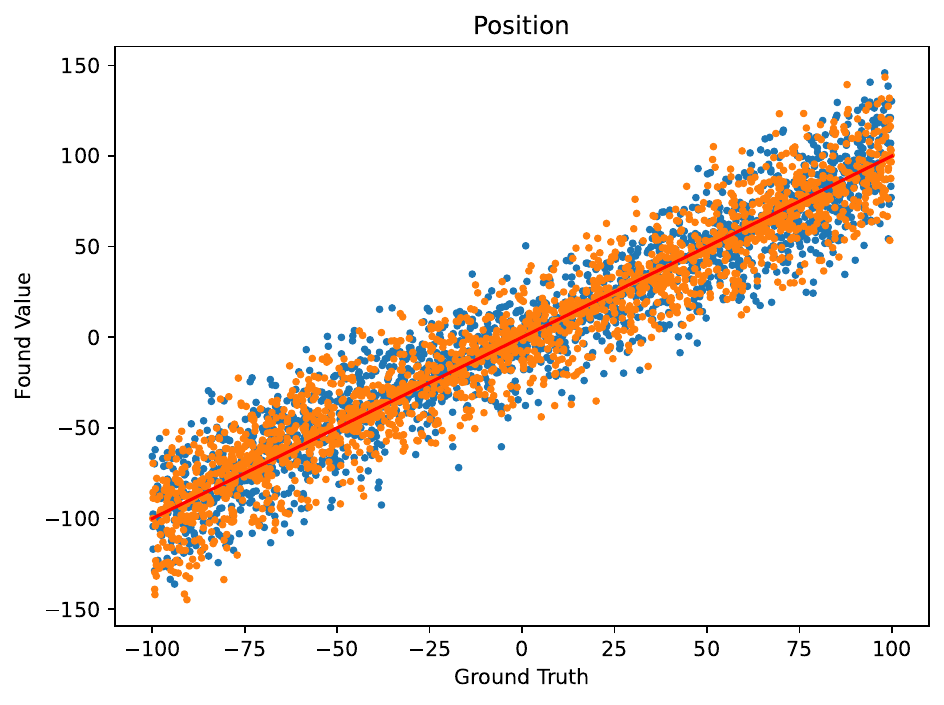}
    \end{subfigure}
\end{figure}

\begin{figure}[p]
    \ContinuedFloat
    \centering
    \begin{center}
        \textbf{Test 2 ($\leq$4th order)}
    \end{center}
    \begin{subfigure}{0.4\textwidth}
        \centering
        \includegraphics[width=\textwidth]{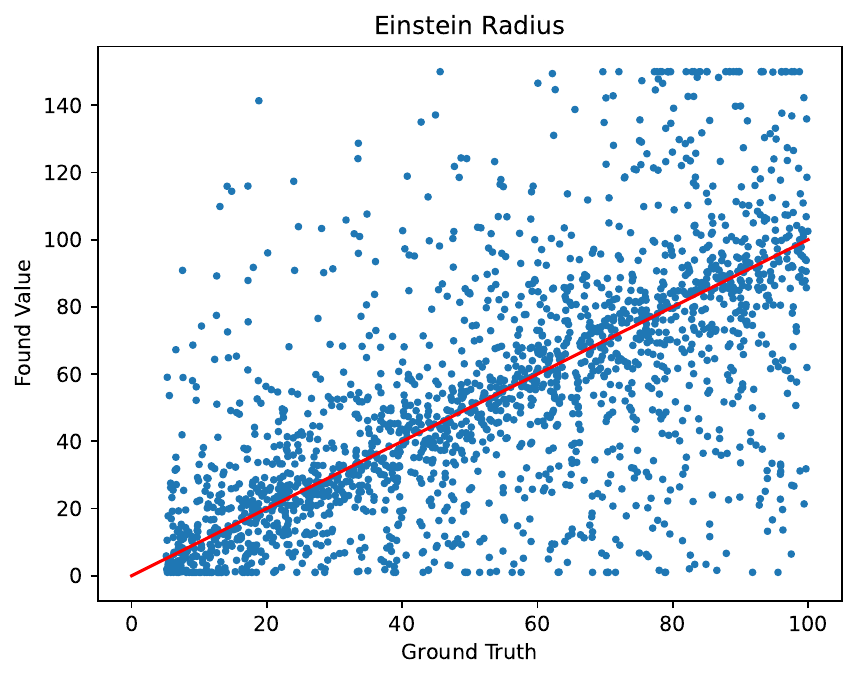}
    \end{subfigure}
    \begin{subfigure}{0.415\textwidth}
        \centering
        \includegraphics[width=\textwidth]{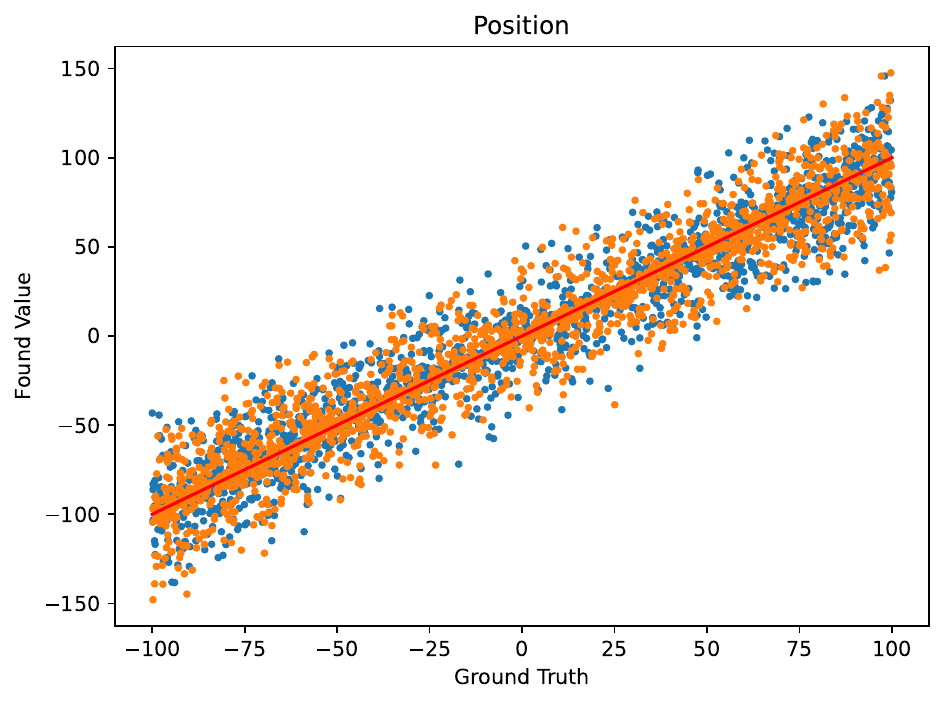}
    \end{subfigure}
\end{figure}

\begin{figure}[p]
    \ContinuedFloat
    \centering
    \begin{center}
        \textbf{Test 3 ($\leq$5th order)}
    \end{center}
    \begin{subfigure}{0.4\textwidth}
        \centering
        \includegraphics[width=\textwidth]{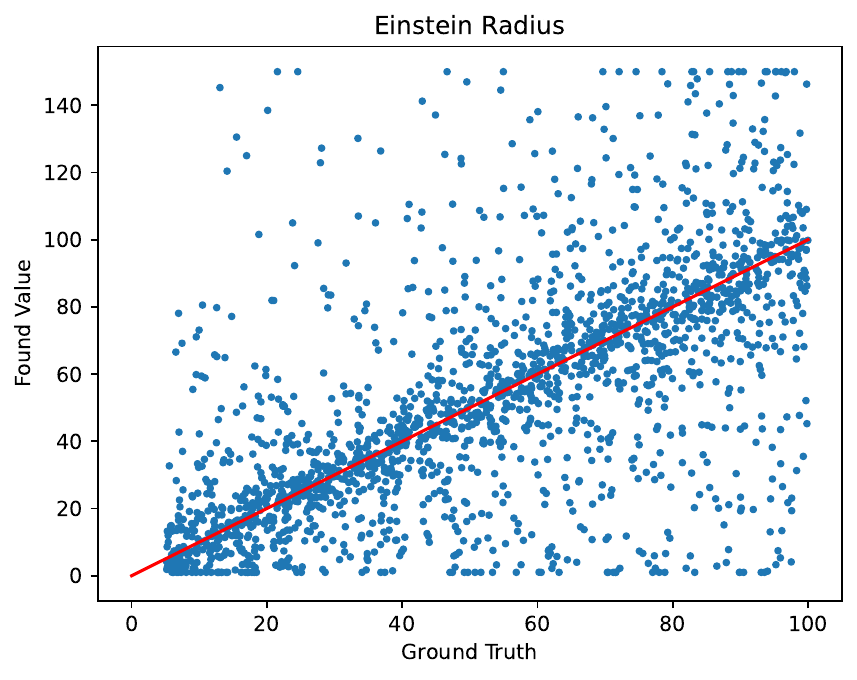}
    \end{subfigure}
    \begin{subfigure}{0.415\textwidth}
        \centering
        \includegraphics[width=\textwidth]{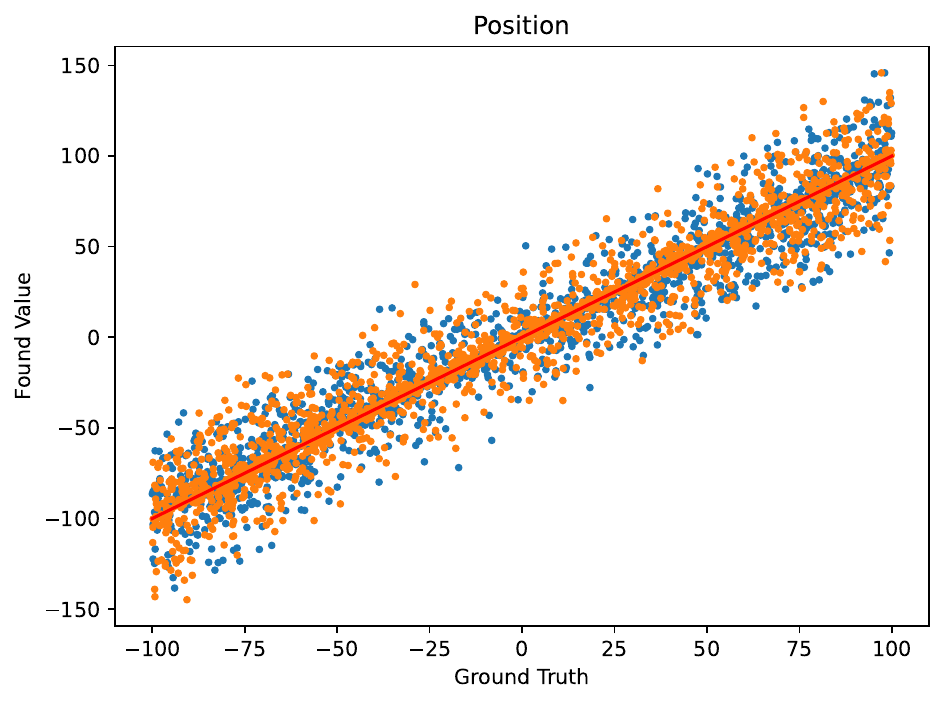}
    \end{subfigure}
    \appfigreset
    \caption{Scenario 3b (SIS cluster lens, with noise): scatter plots of found value versus ground truth for all three tests (rows) and both recovered lens parameters (columns: Einstein radius, position). The red line marks the case where the found value equals the ground truth. For position, the blue (orange) points reference the position in the $\theta_1$ ($\theta_2$) direction. The corresponding $R^2$-values are given in Table~\ref{tab:scenario3b_results}.}
    \label{fig:app_scenario3b_grid}
\end{figure}

%=================================================================
\reftitle{References}
% mdpi.cls already selects \bibliographystyle{Definitions/mdpi} itself
% based on the ACS journal type (astronomy), so we do not set it again
% here (doing so produced a duplicate, illegal \bibstyle command in the
% .aux file and made bibtex fail).
\bibliography{GLbibliography}

\end{document}